# State Vectors and Physical States

## Abstract

Causality and the relativity of simultaneity seem at odds with the apparently sudden, acausal state-vector changes ("collapses") characteristic of quantum phenomena. The problem of how physical phenomena can be causally determined, have the probabilities predicted by quantum theory, and be consistent with special relativity appears to be solved by the assumption, essentially the same as one first used by Aharonov, Bergmann, and Lebowitz to address a different problem, that the "initial" and "final" state vectors of a phenomenon or observation, along with certain other state vectors, *all* represent the system's state *at all times*. Each member of such an aggregate of state vectors is postulated to represent a different aspect of a physical state rather than a state, so that most of the state vectors in effect constitute a set of nonlocal hidden variables. Various implications of this assumption are illustrated through several physical situations, including the "Schrödinger's cat and Wigner's friend" problem, two- and three-particle versions of the Einstein-Podolsky-Rosen experiment, recently devised variations on the double-slit interference experiment, and interacting-particle experiments. Among the results is a somewhat surprising resolution of a paradox in a three-spin Greenberger-Horne-Zeilinger experiment (a "three-particle EPR" experiment), which involves a *logical* consequence that differs from familiar ideas of quantum physics that has no *practical experimental* consequence, and the prediction of new experimental phenomena related to the Zou-Wang-Mandel superposed-idler system. The potential value of the concepts described as heuristics for new predictions and for developing physical intuition by clarifying the interrelation and coherence of physical principles that would otherwise seem contradictory is briefly discussed.

La causalité et la relativité de simultanéité semblent être en désaccord avec les changements, apparemment soudains, de la partie "sans-cause" du vecteur d'état ("effondrements"), caractéristique des phénomènes quantiques. Le problème est de savoir comment la cause d'un phénomène physique, dont les probabilités sont prédites par la théorie quantique peut être déterminée, et également comment elle peut être cohérente avec la relativité restreinte en utilisant une hypothèse. Cette hypothèse est essentiellement la même que celle de Aharonov, Bergmann et Lebowitz, les premiers à l'utiliser pour un tout autre problème, à savoir que : les vecteurs d'état "initiaux" et "finaux" d'un phénomène ou d'une observation, et certains autres vecteurs d'état, tous représentent l'état du système à tout instant. Chaque membre d'un tel agrégat de vecteur d'état est postulé pour représenter un aspect différent d'un état physique plutôt qu'un état même, de sorte que la plupart des vecteurs d'état utilisés constitue un jeu de variables cachées non locales. Différentes implications de cette hypothèse sont illustrées dans plusieurs situations physiques, dont font parti, le chat de Shrödinger et le problème de l'ami de Wigner. La version utilisant 2 et 3 particules de l'expérience d'Einstein, Podolski et Rosen (EPR), et récemment une variante de l'expérience des interférences créées à partir de deux fentes ainsi que les expériences d'interaction de particules, illustrent également cette hypothèse. Parmi les résultats, il y a une solution quelque peu surprenante d'un paradoxe dans l'expérience des 3-spins de Greenberger, Horne et Zeilinger (l'expérience 3-particules d'EPR). Ceci implique une conséquence logique qui diffère des idées familières de la physique quantique, qui n'a pas de conséquence expérimentale pratique, et qui implique également la prédiction de nouveaux phénomènes expérimentaux relatif au système de plusieurs faisceaux laser inactifs superposés de Zou, Wang et Mandel. La valeur potentielle des concepts décrits comme heuristiques pour des nouvelles prédictions est brièvement abordée. Nous discutons également les possibilités pour de nouvelles intuitions physiques en clarifiant la corrélation et la cohérence des principes physiques dans le but d'éviter des contradictions.

# 1.  INTRODUCTION

Of all the conceptual difficulties generally recognized in quantum theory, perhaps the greatest relate to how probabilities are involved.

The characteristics and behavior of real physical systems may be described by state vectors.  A state vector |Ψ> can be represented in terms of other vectors |$v_1$>, |$v_2$>, |$v_3$>, … that together form a basis for the vector space, and in which each $v_i$ stands for a different set of values for some collection of variables v associated with the basis.  Each inner product <$v_i$|Ψ> of |Ψ> with a different |$v_i$> is a complex number, so the generic product <v|Ψ> is a complex function of the set of variables v—what is known as the wave function for |Ψ> in terms of the variable-set v.  Different types of variable v correspond to different representations of |Ψ>; for instance, if the variables are momenta **p**, we have |Ψ> in the momentum representation, and if the variables are sets of position coordinates **q**, we have |Ψ> in the position representation.  If |Ψ> is a state vector for a "one-particle" system, its wave function in the position representation **<q|Ψ>** resembles a classical wave in a classical field.

An important difference from a classical field, which relates to the aforementioned difficulty, is that the position-wavefunction's value at a given point and instant represents, not a field strength in the classical sense, but how likely the particle is to be at that place at that time. More exactly, it represents a probability density "amplitude" for the particle's presence. In typical situations, such wavefunctions represent that the particle has some likelihood of being in any one of a number of places, but no certainty of being in any particular place. On the other hand, a wavefunction that represents a particle that is definitely within some *finite* region of space at a particular instant is of a different type: it will be large within that region and zero elsewhere.

If we determine the locations of particles whose wavefunctions are of the former type, we describe them thereafter with wavefunctions of the latter type. Now any state vector can be represented as a wavefunction whose amplitude at any point (x, y, z) at time t depends causally on the amplitude at immediately neighboring points in the volume (x ± dx, y ± dy, z ± dz) during the immediate past (from t – dt to t); i.e., the wavefunction satisfies a differential equation of motion. But the apparent change in state vectors that attends location of a particle seems hard to account for by any such causal mechanism.

One candidate for a causal mechanism involves the fact that the particle interacts with the measuring apparatus. If the particle assumes a location upon measurement that depends on the measuring apparatus' own initial state, then we should be able to predict that location from the way that a state vector for the combined system particle-plus-measuring-apparatus develops in time. As it happens, though, this is not the case. The equations for how the combined system's state vector develops in time show that both the particle position and the measuring instrument's final reading are uncertain.[1] So if the measuring instrument causes the particle to take up a definite position, that position is determined by something other than the original form of this state vector and the equations for its time development. Thus even the wavefunction amplitudes found for the combined system before and after the measurement lack the kind of causal connection we were looking for.

A state vector evidently describes something real about a system's time development that can be traced experimentally. Nonetheless, fractional probabilities represented by the state vector change to integral probabilities—certainties—when things such as actual particle positions are checked. If in fact no causal mechanism can account for such changes in the state vector, it would seem that the system itself must change acausally.

This apparent acausality has at least three puzzling features.

First, the change itself seems utterly capricious. Indeed, the original state vector could not represent general probabilities unless chance were somehow involved in the system's behavior. But of all the changes that could occur, there is no clear reason why any particular one of those changes should occur rather than one of the other possibilities.

Second, there is the indefiniteness of when the change actually takes place. To take our example further: suppose the particle's position is determined by illuminating it and observing the scattered photon. The equations for the particle-plus-photon system's time development only imply a definite position (and a corresponding state vector) for the particle if we know something definite about the subsequent location of the scattered photon. However, for such a combined system, if the particle's location is uncertain, the original state vector implies that where the photon will be found is correspondingly uncertain, as we discussed above. So presumably one has to somehow illuminate or detect the photon itself. But this wouldn't help: the time-development equation for a state vector that also accounts for the photon detector implies that the photon detector's final condition is just as indefinite as the photon's, inasmuch as the original particle's position is indefinite to start with. Thus state vectors for systems that include chains of detectors and detectors of detectors are no more or less indefinite than those for systems with no detectors. Yet we observe that systems with parameters that can't be evaluated from initially-known state vectors are found to have definite values for those parameters upon interaction with other systems. The question is, when do the parameters become definite? Upon interaction with the next system in the chain? Upon that system's interaction with another system? Judging from the initial state vector of a particle-plus-detector system, it doesn't seem that a particle's position could ever be found. Yet it can be.[1,(1)]

Third, and perhaps most seriously, such changes are apparently instantaneous, or practically so, throughout spatially extended systems. Any sort of change that occurs simultaneously throughout all space in one reference frame will not be simultaneous throughout space in any other reference frame; instead, the change would progress along a moving surface in other reference frames. Yet it always seems as though state-vector changes are instantaneous in whatever reference frame the observer happens to be using. We do not describe the change in state vectors as being progressive throughout space; in certain cases, such as when a particle of well-defined momentum (that is, low momentum-uncertainty) is measured and found within a well-defined region of space, an assumption that the previously-known state vector changed to the other one at the surface of a plane progressing through space would imply that certain probabilities--e.g., the probability for the particle to be at some specific region in space or to have some specific

momentum--would not add or integrate to unity during the state-vector change, if both state vectors were normal.

The first puzzle has received much attention. Different proposals to resolve it have been distinguished by at least two types of underlying postulates: that physical effects are not uniquely determined by physical causes, or that a physical system's state is not completely represented by a state vector. Theories that assume physical causes for quantum systems' behavior in addition to those represented by a state vector would seem to automatically resolve the second puzzle as well.

It has been proven that, if a state vector does not completely describe all physically significant variables of a system, the complete set of variables must connect system behavior at any location with the behavior of whatever might simultaneously measure the system at remote locations.[2] It has been thought that this would mean measuring apparati can propagate effects throughout a system instantaneously.[3] If so, such theories would appear, at least at first sight, to still present the third type of puzzle--perhaps more problematically, since physical influences would definitely be propagated instantly, and not just probability amplitudes which, after all, might be mere abstractions. This question remains even for measurements that require a finite amount of time. The problem is that even the results of those local measurements that require a long time are correlated with the settings of measuring devices too distant to have either influenced them or been influenced by them through any signals traveling at the speed of light or less.

This paper presents an interpretation of quantum physics that appears to resolve all three puzzles. Instead of assuming that one state vector describes the system for one time interval before a measurement, and a different state vector describes the system during another time interval after the measurement, so that the state vector somehow changed with the measurement, the basic assumption is this: *that both state vectors--the one known before the measurement, and the one known after, because of the measurement--together describe the system at all times*. That is, not only is the originally-known state vector assumed to describe the system before a measurement, but during that same time, the other state vector found by measurement is also assumed to already describe that same system, even before the measurement. Likewise, the originally-known state vector is assumed to still describe the same system after the measurement. The two state vectors are thus both assumed to describe different features of the same system concurrently, both before and after the measurement. According to this point of view, then, measurement does not fundamentally change the state vector-- indeed, there is no such thing as "the" state vector. Instead, the implication is that different measurements reveal different state vectors, each of which describes a different already-existing feature of the system.

In saying that each of the state vectors describes some feature of the system both before and after a measurement, it is not implied that the state vectors are not changed in any fashion, but only that measurement is assumed not to change state vectors by "collapse" or any similar mechanism. Since measurements involve interaction between a system and a measuring device or probe, the system is altered by the measurement as the measuring device is. This kind of change is assumed to occur in the state vectors as part of the ordinary time development of the overall "observed-system-plus-measuring-probe" system. (Such time developments, in which the system's present condition determines the condition that immediately follows, we will describe as "causal" time developments.) *What is explicitly not assumed is that measurements do anything to change one state vector to a different one that is not related to it through the system's equation of motion*.

As any measurement might be made on a system, *it is further assumed that the system's complete description includes state vectors whose eigenvalues correspond to each possible kind of measurement and are mutually consistent*. For instance, such a description could include eigenvectors of the position operator whose position eigenvalues were all alike, other eigenvectors of the momentum operator whose momentum eigenvalues were identical, eigenvectors for spin component along the z-axis (some of which would also be eigenvectors of the position operators while others would be eigenvectors of the momentum operators) each of which would have the same eigenvalue (+1 or –1, depending on the system), other eigenvectors of orbital angular momentum operators, and so on. A system would thus be described by a family of state vectors instead of just one. For example, one of the state vectors of this family would correspond to the quantum-theoretical "complete set" of measurements "x-coordinate, y-coordinate, z-coordinate, x-component of spin" for some instant, with appropriate time development for earlier and later instants. This state vector would be one of the infinitely many eigenvectors of this quadruple set of operators, but would be the only eigenvector of this set that actually corresponded to this system. This eigenvector's identity--that is, which eigenvector of these four operators was actually part of the family of state vectors that describe the system--could be revealed by suitable measurements.

Although a system's state is assumed to be represented by a family of state vectors in the interpretation presented here, the elements of such a family are not considered to be components of a "superposition" constituting a single state vector, as one often deals with in quantum theory. While one does, in this interpretation just as in others, deal with state vector superpositions, and represent individual state vectors by component sets in different bases, the nature of the family of state vectors described above is distinct from this. The members of such a family are not the different components of any one state vector in some particular basis; rather, each member is supposed to be the eigenvector of a specific set of operators that describes what one would learn about a particular system by measuring the parameters those operators represented.

The idea that different state vectors represent the same system concurrently over a noninfinitesimal period of time, is essentially the same as one presented in 1964 by Yakir Aharonov, Peter Bergmann, and Joel Lebowitz[4] to address a different problem (namely, whether the time asymmetry associated with the apparent change of a system's state vector upon measurement were intrinsic to the laws of quantum physics or due to the universe's macroscopic irreversibility). Here, we apply the idea to the paradoxical features of quantum theory mentioned above.

How this interpretation resolves the puzzles discussed above is illustrated through a number of examples in the following sections. In one respect, these examples offer no new results. This is perhaps to be expected. Because quantum theory agrees as well as it does with experiment, and because it is not clear that its basic concepts are internally inconsistent, despite their apparent inconsistency with certain other physical concepts, quantum theory appears to be sound. What is new here is an illustration of how the phenomena predicted by the theory can be consistent with causality and relativity.

In each example, one assumption--that the current state of a system corresponds at every instant to one state vector, and that as the state develops in time the same state vector continues to represent it, with the vector's time development following the system's equation of motion--is replaced with the assumption that at any given instant, different aspects of the current state of a system correspond to different state vectors, and that as the state develops in time, the different state vectors continue to represent the same individual aspects of the state as they develop in time, with each vector's time development following the same equation of motion. Or, put more simply, the assumption that state vectors correspond one-to-one with physical states is replaced with the assumption that state vectors correspond one-to-one with single features or aspects of physical states.

By using the term "state" in this way, we are in essence identifying a system's current state with its past and future states and using the same term for both the system's momentary condition and its entire time development. We may do so, since the past, present, and future conditions each imply the others, given the interactions that govern the system. This will also be convenient, since the relationship between the system's (developing) condition and the state vectors used to describe it is one of the main concepts to be examined in this paper, and using the term "state" for "momentary condition and how it develops through time" will let us focus on this relationship without repeatedly referring to inessential points.

Since different paradoxical consequences of the more usual assumption may be most easily recognized in different types of experiments, this paper treats a variety of examples to illustrate the paradoxes' resolution.

Section 2 presents a very simple physical example, in which the state vectors' features of interest remain constant in time. Later sections present other examples to clarify how the three puzzles can be resolved when the originally-known state vector changes with time and the state vector found by later measurement has a distinct time development of its own. Of course, the individual state vectors' own causal time developments are no more paradoxical in quantum theory than in classical theory; the puzzles relate to how systems appear to change from a condition described by the first-known state vector to a condition described by a different vector. When we discuss how state vectors appear to change, we will generally be referring to this sort of apparent change, not to the time development of any individual state vector.

While Section 2 describes a system with one component, some of the better-known illustrations of quantum theory's paradoxes involve systems with many interrelated components. Sections 3 through 5 deal with variations of the Schrödinger's cat/Wigner's friend problem and with the Einstein-Podolsky-Rosen paradox, beginning with an analysis in terms of the basic idea presented in this section, continuing with an examination of these two systems' time development, and ending with a review of analogies between the two systems and features common to all two-component systems. Section 3's review of Bell's theorem also uses the same basic concept to address the problem of "hidden variables" in what appears to be a new way. Section 6 makes use of the preceding sections' analogies in resolving paradoxes of double-slit experiments, with particular emphasis on the version proposed by Scully, Englert, and Walther.[5,6]

Problems raised by a slightly more complicated system that involves three interacting components instead of two are discussed in Section 7. Section 8 shows that for this case, the interpretation of quantum physics presented in Section 2 leads to a logical consequence differing from that of more usual interpretations, though not to different experimental consequences.

For different experimental consequences of this interpretation, Section 9 presents another variation of the double-slit experiment that features some surprising phenomena of its own. This section particularly illustrates how assumptions about the mechanism of such phenomena, though consistent with usual interpretations of quantum theory, may appear contrary to other physical principles. Such assumptions would also imply experimental possibilities contrary to those suggested by the interpretation presented here. Immediately after Section 10's brief review of the points discussed in previous sections and its look ahead to the remaining ones, Section 11 shows what this interpretation implies about certain features of elementary-particle interactions.

Section 12, using the photoelectric effect and the Schrödinger's cat problem as examples, discusses how the way we normally perceive quantum phenomena may relate to usual interpretations of them. Section 13 discusses implications for the question of state vectors for the universe as a whole. Finally, Section 14 discusses the potential value of the new interpretation of quantum physics.

## 2. ILLUSTRATION: SYSTEM OF ONE COMPONENT

Consider a particle whose spin-direction is established to be positive along a particular axis m. The state vector corresponding to this may be denoted |+m⟩. If the particle's spin is measured along any other axis (call it the n-axis), it will be described by a different state vector, either |+n⟩ or |-n⟩. According to standard interpretations of quantum phenomena, this means that the particle's state vector will *change* from |+m⟩ to the appropriate "n-axis" vector. This assumption leaves us with the puzzles mentioned above.

As mentioned in Section 1, an alternative interpretation is to assume that the different state vectors, rather than representing the particle's state at different times, both represent the particle at all times. According to this point of view, when the particle is prepared so that its spin in the m-direction is positive, the spin will also have a particular orientation along the n-axis, and both spin orientations are concurrently described by distinct state vectors. The measured n-orientation will be positive or negative according to whether the already-existing "n" state vector is |+n⟩ or |-n⟩.

This alternative view, in which the system's state vectors are not considered to be related to a change of state, leaves us with none of the puzzles mentioned above. Since in this view the second state vector is revealed by measurement rather than caused by it, nothing about the particle is changed, and the outcome of the measurement is not at all capricious. Since the revealed state vector already exists before the measurement, and in fact coexists with the already-known state vector for spin along the m-axis, it cannot be the result of a change in the originally known state vector; there is thus no indefiniteness in when a change from one state vector to the other happens. And for the same reason, the problem of instantaneous change in the particle's spin-direction does not exist, even if

State Vectors and Physical States    18

neither state vector indicates a definite location for the particle; there is no change from the |+m⟩ to either the |+n⟩ or the |-n⟩ state vector, since the m-axis and n-axis vectors both describe the particle everywhere at all times, so there is no question of whether the change is instantaneous, or why the change, if instantaneous, is so in one frame of reference rather than another.

If this is true, the measurement one finds is determined by something other than the measuring interaction. The aforementioned probabilities of quantum physics, according to this interpretation, are not probabilities for a physical system to change from one state with its time development to another state with a different time development, but are probabilities that a system's state, if characterized by one state vector, happens to be equally well characterized by another state vector. In other words, state vectors are not equated with states; any given physical state is associated with multiple state vectors, only one of which can be determined by any single measurement, but all of which together completely characterize the physical state.

While this illustration addressed all the paradoxical features mentioned in Section 1, the difficulties are perhaps better known through consideration of more familiar examples involving slightly more complex systems. The puzzle from more familiar interpretations of when the change in state vector or wavefunction occurs is well known from the situation illustrated by Schrödinger's cat and Wigner's friend, while the problem of simultaneity, or apparent "action at a distance", is perhaps most widely recognized from consideration of the Einstein-Podolsky-Rosen experiment and variations thereon. We will examine these experiments in the next few sections.

# 3. PARADOX RESOLUTIONS THROUGH MULTIPLICITY OF STATE VECTORS: TWO-COMPONENT SYSTEMS

Let us first reexamine the problems of when a quantum system's state changes and what causes the change in more detail, and consider how the idea of concurrent state vectors resolves them. In this section, we illustrate these considerations with two thought experiments: the first being an experiment related to the Schrödinger's cat and Wigner's friend problems[2,(7,8)], and the second being John Bell's variant of the Einstein-Podolsky-Rosen (or EPR) experiment which he used in his analysis of nonlocal hidden variables in quantum physics.[(2)]

To avoid complications from the additional questions related to whether Schrödinger's cat is alive, dead, or in some sort of intermediate state, and when, let us reconsider our inanimate particle with spin. Assume that our particle is initially in a state such that its spin in the x-direction is positive, but that it passes through a device which is set to either reorient this spin to the negative x-direction or leave it alone, according to whether a hollow, hermetically sealed sensor attached to the device detects, within a given set time interval, an alpha particle from a radioactive atom contained inside it, the interval being chosen so that the detection probability is 1/2. The device could affect the particle's spin orientation by a suitably changing magnetic field: while an unchanging magnetic field along the x-axis would leave the particle's spin positive in that direction, a field whose orientation was rotated to the opposite direction while the particle passed through it would, with nearly 100% probability for sufficiently slow rotation, reorient the particle's spin the same way. The field would be rotated or left as it was depending on whether or not the radioactive atom disintegrated in the appropriate interval.

If this spin were the ordinary (essentially "orbital") angular momentum of a classical rotating body, that angular momentum would either have its original orientation or the opposite one, and no components along any other direction, once the body had crossed the device. For quantum-physical spin angular momentum, the results of crossing the device are different. First, once the particle has passed through the spin-orienter, the particle/spin-orienter system is properly characterized by a state vector that is a linear combination of two vectors--one describing the particle as having a spin in the positive x-direction and the radioactive atom as undecayed, and the other describing the particle's spin as being oriented in the negative x-direction while the atom in the sensor is decayed. Nevertheless, the spin will still be found to have a definite orientation along the x-axis, *if* we look for that orientation. We could, for instance, have the particle leave the spin-orienter and pass through a Stern-Gerlach apparatus aligned with the x-axis, see which direction the particle is deflected into, and thereby determine the spin's orientation along that axis. And this orientation will correlate exactly with the state of the radioactive source--negative if the source is found to have indeed radiated an alpha particle into the detector, positive if not. In conventional interpretations, this is taken to mean that the particle has somehow changed its state, from one characterized by lack of a definite x-spin to one with either a positive or negative x-spin, as the case may be. Aside from the apparent lack of causation for this, there is the question of when such change actually occurs. No state vector other than the one that develops from the originally-observed one according to the quantum-physical equation of motion is evident unless behavior represented by a different state vector is checked for, and conventional interpretations hold, or at least suggest, that the observation has something to do with making the change happen.

But if observation causes the change, what observation causes it? Determining the particle's path from the Stern-Gerlach apparatus requires the use of some kind of

State Vectors and Physical States    21

sensor. We could, for instance, put sensors on the apparatus' exit end and connect them with an indicator that produces one signal when the particle is found leaving in the positive x-direction and a different signal when the particle is found leaving in the negative x-direction. The state vector for this augmented system--which includes not only the particle plus the device that either does or doesn't change its spin's x-orientation, but the Stern-Gerlach apparatus that deflects the particle along that x-orientation, *and* the sensor that reacts differently in either case--will itself not develop into one for which the spin is either "+x" or "-x", but again into one that is mathematically a linear combination of the two. Yet if the new sensor set is checked to see whether its state corresponds to "+x" or "-x", it will be found to definitely correspond to one or the other, with a 50% probability for either. Checking whether the sensor's state corresponds to "+x" or "-x" requires another sensor, to react to the signal from the first sensor. Yet by the same reasoning, the state vector for the system that includes this other sensor will be a linear combination of state vectors that each represent just one of the possibilities, not just one of these vectors or the other. But again, with proper types of sensors, we always do find one orientation or the other--half the time "+x", half the time "-x", but always just one of them. So when does the spin's change from an indefinite and undefinable x-orientation to a definite one occur? When the particle's path is detected by the first sensor? Or when that sensor's signal affects a second sensor?

Continuing this reasoning, we can determine that no matter how long a chain of sensors we set up, there is no place at which any physical cause reflected in the system's originally-considered state vector, or in its equation of motion, can make any sensor definitely indicate whether the particle's spin ended up in either the positive or the negative x-direction. Yet any sensor along the way, if checked, will be found to indicate one direction or the other, and its reading will be consistent with that of every other sensor in the chain. If the state vector observed has any relation to the system's physical state, when does it change? And why does it change at that particular stage rather than at one of the other earlier or later ones?

No such problem arises from the assumption that a physical system's state is generally characterized by more than one state vector. If we assume that the combined system of a particle with spin and a radioactive atom is originally characterized, not only by a state vector corresponding to positive x-orientation for the spin and existence of an intact, unstable nucleus in the atom, but by another state vector as well, whose final form corresponds either to positive x-spin and intact nucleus or to negative x-spin and decayed nucleus, one or the other, then the system stays in a single state throughout--a state that can only be characterized by both state vectors (and others as well, as we shall see), each of which *independently* develops in time according to the same equation of motion.

From the initial form of the particle-with-spin/radioactive-nucleus system's initially-known state vector, the latter form--the one indicating a 50% probability for each possible x-orientation of the spin and corresponding state of the radioactive nucleus--can be readily calculated by the usual methods of quantum theory. The exact initial form of the last-found state vector for the particle-with-spin/radioactive-nucleus system--the one that corresponds to the definite x-orientation of the particle's spin after the spin was or was not reoriented--can be calculated likewise. However, the significance of this state vector's initial form might best be understood after consideration of Bell's version of the EPR experiment, which involves a system whose constituents are similarly interrelated, but which resemble each other more and interact more directly than the particle with spin and the radioactive nucleus.

Bell considered pairs of particles with spin, each pair of which is a system characterizable by a "singlet-state" vector, that originally interact but later become separated by a mechanism that leaves the particles' individual spins unaffected. The orientations of each pair's spins along any axes could be measured individually at any time after the separation. For each pair, if the axes are different, neither spin necessarily correlates with the other, though the frequencies of particular pairs of orientations match the probabilities expected from quantum theory. But if the axes for both particles are the same, the orientations are always opposite, again in accordance with quantum theory.

It has been supposed that the spins one finds for each particle of a pair in such experiments might be predetermined by a set of variables that includes more than the pair's state vector. Bell's analysis shows that if this is true, then whether the orientation of each particle's spin is parallel or antiparallel to a given axis must depend not only on what axis this is, but on which axis the other particle's orientation is measured along too-- no matter where or how far away the other particle's spin happens to be measured from the first's. Bell took this result to mean that, if variables beyond those of the state vector did determine the spin orientations found along particular axes, information about which axis either particle's spin was being measured along would have to be instantly propagated to the other particle, so that the probabilities of the other particle's spin being parallel or antiparallel to whatever axis *it* was measured along would be the quantum-theoretical ones.

The idea that the state of any physical system is characterized not by one state vector, but by a set of many concurrent state vectors, suggests that instantaneous propagation of this sort need not occur. The review of Bell's argument below shows that, if more than one state vector does characterize a physical system such as the "singlet-state" spin-pair, one of the "other" state vectors will in essence constitute the set of additional variables necessary to determine the spins' orientation along whatever axes they will be measured against. In other words, each of these other state vectors for the actual system would determine the spin-orientations that would be observed along one possible pair of measurement-apparatus axes, without instantaneous information transfer between measurement apparati.

Bell considered the expectation value P(**a**, **b**), defined as the product of the spin-orientations of particles 1 and 2 along the directions of the respective unit vectors **a** and **b**, each orientation being +1 if parallel to its respective axis and -1 if antiparallel. The two orientation variables, symbolized by A and B respectively, would depend on the unit vectors **a** and **b** and on the set of values λ of all the variables that characterize the two-particle system itself--which might be a more comprehensive set than those specifying the "singlet-state" state vector (in Bell's argument, actually the wave function corresponding to that state vector). If ρ(λ) is the probability distribution of the variable values λ, then

$$P(a,b) = \int \rho(\lambda) AB d\lambda .\qquad (1)$$

(Here the parameter set λ is treated as continuous; corresponding expressions involving summation rather than integration could represent cases in which λ includes discrete variables.) Bell went on to prove that, if A is independent of **b** and B independent of **a**, the expectation value P(**a**, **b**) neither equals the value calculated from quantum mechanics, -**a**·**b**, nor approximates it arbitrarily closely.

If we assume that the system of particles described has one state vector for every possible pair of directions **a** and **b** along which the spin-orientation meters can be aligned, we find that P(**a**, **b**) does equal -**a**·**b**. Let the "singlet-state" state vector be represented by |Ψ>, and the state vector corresponding to the pair of alignments A and B be represented by |A, B>, so that |+1, +1> stands for the state vector corresponding to

both spins A and B being parallel to the unit vectors **a** and **b** respectively, |+1, -1> stands for the similar vector for particle 1's spin being parallel to **a** while particle 2's spin is antiparallel to **b**, &*c*. Note that |A, B> is an eigenvector of the operators $\sigma_1 \cdot \mathbf{a}$ and $\sigma_2 \cdot \mathbf{b}$, in which $\sigma_1$ and $\sigma_2$ represent the spins for particles 1 and 2 respectively, and that A and B are the corresponding eigenvalues. Since each $\sigma$ is shorthand for $\sigma_x \mathbf{x} + \sigma_y \mathbf{y} + \sigma_z \mathbf{z}$, $\sigma_x$, $\sigma_y$, and $\sigma_z$ being the usual set of 2x2 spin matrices in some representation[3,(9)], $\sigma_1 \cdot \mathbf{a}$ and $\sigma_2 \cdot \mathbf{b}$ can be represented by the matrices $\sigma_{x1} a_x + \sigma_{y1} a_y + \sigma_{z1} a_z$ and $\sigma_{x2} b_x + \sigma_{y2} b_y + \sigma_{z2} b_z$. If $\sigma_x$, $\sigma_y$, and $\sigma_z$ are represented by $[[0, +1]^T, [+1, 0]^T]$, $[[0, +i]^T, [-i, 0]^T]$, and $[[+1, 0]^T, [0, +1]^T]$, respectively, $\sigma_1 \cdot \mathbf{a}$ and $\sigma_2 \cdot \mathbf{b}$ can be represented by $[[+a_z, +a_x+ia_y]^T, [+a_x-ia_y, -a_z]^T]_1$ and $[[+b_z, +b_x+ib_y]^T, [+b_x-ib_y, -b_z]^T]_2$, while |A, B> can be represented by pairs of vectors selected appropriately from among $(2(1 \pm a_z))^{-1/2} [a_z \pm 1, a_x + ia_y]^T{}_1$ and $(2(1 \pm b_z))^{-1/2} [b_z \pm 1, b_x + ib_y]^T{}_2$, the sign choices depending on the eigenvalues A and B.

If |Ψ> and the variable state vector |A, B> are the parameters in the set λ that P(**a**, **b**) actually depends on, P(**a**, **b**) is a sum rather than an integral since |A, B> is discrete:

$$P(\mathbf{a},\mathbf{b}) = \sum_\lambda \rho(\lambda) AB. \qquad (2)$$

A and B are both functions of the elements of λ, and can be represented as <A, B|$\sigma_1 \cdot \mathbf{a}$|A, B> and <A, B|$\sigma_2 \cdot \mathbf{b}$|A, B>, respectively. ρ(λ), the probability that the "singlet-state" system will be found to have spin-orientations A and B along the unit vectors **a** and **b**, equals |<A, B|Ψ>|². Thus

$$P(\boldsymbol{a},\boldsymbol{b}) = \sum_\lambda \left|\langle A,B|\Psi\rangle\right|^2 \langle A,B|\sigma_1\cdot\boldsymbol{a}|A,B\rangle\langle A,B|\sigma_2\cdot\boldsymbol{b}|A,B\rangle$$

$$= \sum_\lambda \langle\Psi|A,B\rangle\langle A,B|\sigma_1\cdot\boldsymbol{a}|A,B\rangle\langle A,B|\sigma_2\cdot\boldsymbol{b}|A,B\rangle\langle A,B|\Psi\rangle. \qquad (3)$$

This is equal to the quantum-mechanical expectation value of $(\sigma_1\cdot\boldsymbol{a})(\sigma_2\cdot\boldsymbol{b})$, as is evident from the following. First,

$$P(\boldsymbol{a},\boldsymbol{b}) = \sum_\lambda \langle\Psi|A,B\rangle\langle A,B|\sigma_1\cdot\boldsymbol{a}|A,B\rangle\langle A,B|\sigma_2\cdot\boldsymbol{b}|A,B\rangle\langle A,B|\Psi\rangle$$

$$= \sum_\lambda \langle\Psi|A,B\rangle\langle A,B|A|A,B\rangle\langle A,B|B|A,B\rangle\langle A,B|\Psi\rangle$$

$$= \sum_\lambda \langle\Psi|A,B\rangle\langle A,B|A,B\rangle\langle A,B|A,B\rangle\langle A,B|\Psi\rangle AB \qquad (4)$$

$$= \sum_\lambda\sum_\mu\sum_\nu \langle\Psi|C,D\rangle\langle C,D|A,B\rangle\langle A,B|E,F\rangle\langle E,F|\Psi\rangle AB.$$

(Here, μ and ν are essentially alternative expressions for |C, D> (or <C, D|) and |E, F> (<E, F|), respectively, just as the sum over different λ amounts to a summation over different values of |A, B> (or--equivalently here--of <A, B|), all of these variable expressions representing possible state vectors for spin-orientations of particles 1 and 2 along the respective axes **a** and **b**. The triple sum, which includes the products of all possible combinations of |A, B><A, B|, |C, D><C, D|, and |E, F><E, F|, is equal to the sum preceding it since the product <C, D|A, B><A, B|E, F> is zero unless C=A=E *and* D=B=F.)

Continuing,

$$P(a,b) = \sum_\lambda \sum_\mu \sum_\nu \langle \Psi | C,D \rangle \langle C,D | A,B \rangle \langle A,B | E,F \rangle \langle E,F | \Psi \rangle AB$$

$$= \sum_\lambda \sum_\mu \sum_\nu \langle \Psi | C,D \rangle \langle C,D | A | A,B \rangle \langle A,B | B | E,F \rangle \langle E,F | \Psi \rangle$$

$$= \sum_\lambda \sum_\mu \sum_\nu \langle \Psi | C,D \rangle \langle C,D | \sigma_1 \cdot a | A,B \rangle \langle A,B | \sigma_2 \cdot b | E,F \rangle \langle E,F | \Psi \rangle \quad (5)$$

$$= \sum_\mu \sum_\nu \langle \Psi | C,D \rangle \langle C,D | (\sigma_1 \cdot a)(\sigma_2 \cdot b) | E,F \rangle \langle C,D | \Psi \rangle$$

$$= \langle \Psi | (\sigma_1 \cdot a)(\sigma_2 \cdot b) | \Psi \rangle,$$

which equals -**a**·**b**, the expectation value of ($\sigma_1$·**a**)($\sigma_2$·**b**) according to quantum theory.

Bell's conclusion was that, if variables other than the wavefunction were to predetermine the outcomes of measurements, those measurements' statistics could only be consistent with those predicted by quantum theory if some mechanism existed whereby one measuring device's setting could influence the reading of another instrument, however remote. Now in the demonstration above, we considered functions A and B that indeed both depend on the orientations **a** and **b** of both measuring devices. However, the idea of this physical process is quite different from the one discussed by Bell. Rather than assuming that information about either spin-measuring device's orientation is transmitted instantaneously to the other device, the set of actual particle spin-orientations implied by the actual state vectors--one state vector for each operator ($\sigma_1$·**a**)($\sigma_2$·**b**)--is assumed to be a preexisting trait of the two-particle system. Furthermore, as is shown below, every type of state vector that can represent some aspect of this system may be assumed to exist already, each type corresponding to some set of measurable features of the system.

Thus, instead of assuming a one-to-one equivalence between state vectors and states of a physical system, it is assumed that a multitude of state vectors is needed to completely characterize one state. As was suggested in the discussion of the single-spin system, the probabilities that characterize quantum theory are accordingly assumed to be probabilities that, if a given state vector |K> describes some feature k of the system's state, another given vector |L> describes a corresponding feature l of that same state. In Bell's terms, the set of all "other" state vectors |L> is the set of nonlocal "hidden variables" that, together with |K>, determine what will be observed when any one aspect of the system is measured.

For this scheme to be consistent, it turns out that not just any state vectors can go together to describe a particular state. Our usual quantum-physical probability calculations involve considering state vectors two at a time. But if measurements taken on one part of a system are to be independent of what quantities of a remote part of the system are measured, with only the state vectors of the system itself determining what we will find, that entire set of state vectors has to be self-consistent.

In our "two-spin singlet-state" example, it is easy to see that self-consistent subsets exist that specify spin-orientations for both particles along the same axis, since the two particles need only to have opposite spins along that axis. There are only two possible choices for each axis: either particle 1's spin is positive along that axis (while particle 2's is negative), or particle 1's spin is negative along that axis (while 2's is positive). But other state vectors should exist as well, which represent what would be found if the spin-orientations of the two particles were measured along different axes rather than the same axis. For the entire set of state vectors to be consistent, whatever spin-orientations are specified for each particle by the "different-axes" state vectors need only be the same as those specified by the "same-axis" vectors. The idea here is that a measurement of the spin of particle 1 is a measurement of what any and all state eigenvectors of that spin determine about it. Indeed, any two state vectors that specify different values for any one quantity are orthogonal, and the probability that they both represent the same state of the same system is zero. The probability that any possible pair of spin-orientations along arbitrary axes (A, B) is consistent with the singlet state is calculable in the usual manner (and is represented by $|<\Psi|A, B>|^2$). The probability that any pair (A', B') of orientations--along the same or other arbitrary axes--is consistent with the pair (A, B) is similarly calculable. Note, however, that the latter probability is unrelated to whether or not the pairs in question pertain to a system in the singlet state. The probabilities $|<A', B'|A, B>|^2$ would be the same regardless of whether or not the singlet state vector $|\Psi>$ characterized the system, being conditional probabilities independent of this.

The consistency of single individual pairs of vectors is assumed in customary interpretations of quantum theory, and is thus not particularly surprising, but as is shown below, this consistency of entire sets of state vectors has interesting implications for slightly more complicated systems. Before this demonstration, though, there are some further things to consider about two-constituent systems.

## 4. TIME-DEPENDENCE OF INDIVIDUAL STATE VECTORS

The notation used above for the state vectors considered here, $|\Psi>$ and $|A, B>$, does not explicitly refer to any dependence on time, since for the preceding mathematical analysis, what matters is the unchanging character of the spin-orientations associated with each vector. In practice, though, features of a physical system are measured at particular places and times. Information on where these features are measured and when, along with what they imply in conjunction with the quantum-theoretical equation of motion or the related S-operators, can be included along with the spin-character notation to indicate each observed state vector's time-dependence and thus specify the vectors more fully. In this section we look at the time development of both the state vectors we dealt with when considering the Schrödinger's cat/Wigner's friend problem, and note an important implication of how the originally-known state vector develops after the particle crosses the spin-orienter and what the finally-observed state vector was like before then.

In the experiment we examined to consider the Schrödinger's cat/Wigner's friend problem, we had one spin and a radionucleus instead of two spins. Because of how this system's constituents interact, the state vector for the system that tells us the particle has a spin initially oriented along the positive x-direction does not define this spin once the particle has crossed the spin-orienting apparatus. Thus the state vector in this case is time-dependent. But the spin component we measure after the particle has interacted with the spin-orienter is along the same axis, and has a definite orientation. Furthermore, we can observe that if this spin is in the positive direction, the nucleus in our apparatus didn't decay in the specified time interval, and if the direction is negative, the nucleus did decay then. The possible state vectors *for the x-direction of the spin and the integrity of the nucleus* are thus

$$|+x, \text{still-intact radionucleus}\rangle$$

and

$$|-x, \text{decayed radionucleus}\rangle.$$

How the system's state vectors change from one instant to the next could be represented by a Hamiltonian operator. For this presentation's immediate purposes, though, it will suffice to consider the S-operator that relates the form of each state vector before the particle and spin-orienter begin interacting to the form of the same vectors after the interaction is over.

If the radioactive nucleus is not found to decay within the set time interval, a state vector of the form "|+x, still-intact radionucleus>" will be found to describe the system just as well at the end of the experiment as at the beginning. Thus one term of the S-operator must be proportional to the identity operator. Alternatively, "|-x, decayed radionucleus>" is a possible final state vector. If the operator $a_X$ is such that

$$a_X|+x\rangle = |-x\rangle, \qquad (6)$$

and the operator $b_X^\dagger$ is such that

$$b_X^\dagger|\text{still-intact radionucleus}\rangle = |\text{decayed radionucleus}\rangle, \qquad (7)$$

then the S-operator for this system also has a term proportional to $a_X b_X^\dagger$. Furthermore, there must also be a term proportional to $a_X^\dagger b_X^\dagger$, to account for what happens when the nucleus decays and the spin was antiparallel to the x-axis to begin with. Finally, since S-operators are Hermitian, this S-operator also must include terms proportional to $a_X^\dagger b_X$ and $a_X b_X$, the Hermitian conjugates of $a_X b_X^\dagger$ and $a_X^\dagger b_X^\dagger$. From these considerations, and in view of state-vector normalization and the fact that this operator and its Hermitian conjugate must be inverses, we find that the S-operator for this system can be represented by

$$(1 + e^{+i\phi}a_X b_X^\dagger + e^{+i\chi}a_X^\dagger b_X^\dagger - e^{-i\phi}a_X^\dagger b_X - e^{-i\chi}a_X b_X)/\sqrt{2} \tag{8}$$

(in which $\phi$ and $\chi$ are arbitrarily defined phases), provided that

$$a_X^\dagger |+x\rangle = 0 \tag{9}$$

and

$$b_X |\text{intact radionucleus}\rangle = 0. \tag{10}$$

Thus the state vector whose initial form is "$|+x, \text{still-intact radionucleus}\rangle$" is found to have the final form

$$\begin{aligned} S|+x, &\text{still-intact radionucleus}\rangle \\ &= (|+x, \text{still-intact radionucleus}\rangle \\ &\quad + e^{+i\phi}|-x, \text{decayed radionucleus}\rangle)/\sqrt{2}. \end{aligned} \tag{11}$$

Likewise, each of the state vectors we might see at the end of the experiment when we measure the particle's spin in the x-direction and examine the integrity of the radionucleus, represented by $\langle +x, \text{still-intact radionucleus}|$ and $\langle -x, \text{decayed radionucleus}|$, are found to have the initial forms

$$\begin{aligned} \langle +x, &\text{still-intact radionucleus}|S \\ &= (\langle +x, \text{still-intact radionucleus}| \\ &\quad - \langle -x, \text{decayed radionucleus}|e^{-i\phi})/\sqrt{2} \end{aligned} \tag{12}$$

and

$$\langle -x, \text{decayed radionucleus}|S$$
$$= (\langle -x, \text{decayed radionucleus}| \quad (13)$$
$$+ \langle +x, \text{still-intact radionucleus}|e^{+i\phi})/\sqrt{2},$$

or (using the "ket" notation)

$$(|+x, \text{still-intact radionucleus}\rangle \quad (14)$$
$$- e^{+i\phi}|-x, \text{decayed radionucleus}\rangle)/\sqrt{2}$$

and

$$(|-x, \text{decayed radionucleus}\rangle \quad (15)$$
$$+ e^{-i\phi}|+x, \text{still-intact radionucleus}\rangle)/\sqrt{2}$$

respectively.

This shows that, just as the initially-observed state vector |+x, still-intact radionucleus⟩ develops into one that specifies neither the spin's x-direction nor the radionucleus' integrity, the state vector (whichever one it is) that specifies definite values for both *after* the interaction *originally* specifies neither.

# 5. MATHEMATICAL CORRESPONDENCES AMONG TWO-COMPONENT SYSTEMS

In this section we point out analogies between the particle-radionucleus system and the two-particle system of the EPR experiment, as an aid to better understanding both. In so doing, we will note features common to all experiments with a two-component system. This will prepare the way for discussions of other two-component and multi-component systems in later sections.

Aside from the time-dependence, the particle-radionucleus system is mathematically similar to that of the EPR experiment. Once the particle interacts with the spin-orienter, the state vector that was originally

$$|+x, \text{intact radionucleus}\rangle$$

has the form

$$(|+x, \text{still-intact radionucleus}\rangle + e^{+i\phi}|-x, \text{decayed radionucleus}\rangle)/\sqrt{2}.$$

If we think of a decayed radionucleus as a "+x" (i.e., as one that has "positively" emitted some particle) and a still-intact one as a "-x", and in so doing choose $\pi$ for the phase $\phi$, we can see that this state vector's final form is expressible as "$(|+x, -x\rangle - |-x, +x\rangle)/\sqrt{2}$". Now the singlet-state's state vector $|\Psi\rangle$ considered by Bell, which is mathematically definable by a particular linear combination of definite-spin pairs, can be described in a variety of ways. In the "$|A, B\rangle$" notation, "A" and "B" stood for the eigenvalues of the spin along the directions of unit vectors **a** and **b**. If **a** and **b** are identical but otherwise arbitrary, the state vector |ψ⟩ is expressible by (|+1, -1⟩ - |-1, +1⟩)/√2. If we wished, we could provide more detail in the notation and indicate which spatial axis these eigenvalues refer to, for instance expressing |Ψ⟩ by (|+a, -a⟩ - |-a, +a⟩)/√2 to specifically indicate spin-orientations in the "+a" or "-a" directions. If **a** = **x** (the unit vector in the positive x-direction), we would thus represent |ψ⟩ by "(|+x, -x⟩ - |-x, +x⟩)/√2", which has the same form as the state vector just considered for the particle-nucleus system.

We can thus describe the particle-nucleus system as being in something like the singlet state of the EPR experiment. The end result of the decay or cohesion of the radionucleus that does or does not trigger the reorienting apparatus is analogous to the spin of "EPR particle 2" along a particular axis. Other two-constituent systems, each constituent of which has one of two possible values for certain quantities, are likewise analogous to these, being describable by the same sort of mathematical expressions.

Although these systems are similar, they clarify different paradoxes differently. We considered the particle-nucleus system as we addressed the question of when a state vector changes due to measurement, while the singlet state of a pair of particles with spin brought to our attention the idea that action at a distance seems necessary to account for correlations in separated constituents of a system. But recognizing the mathematical similarities between these systems helps us clarify which phenomena in one system are analogous to which phenomena in another. Awareness of this kind of similarity can help us correctly figure out less-understood features in one system by finding appropriate better-understood features to consider in another.

As an illustration of this, we can see that although neither constituent of the two-particles-with-spin system undergoes a torque to reorient its spin as the one-particle-with-spin/radioactive-nucleus system does, the spin of one constituent of the particle/nucleus system can be measured along any axis, just as the spin of either constituent of the two-particle system can. We may also regard the integrity or decay of the nucleus as analogous to opposite spin-orientations along an axis, with nuclear conditions represented by state vectors like "(|intact> ± $e^{+i\phi}$|decayed>)/√2" being analogs of spin-state vectors like "(|-x> ± $e^{+i\phi}$|+x>)/√2", which themselves correspond to positive or negative spin-orientations along different axes in the y-z plane. We may thus more easily understand that the two-spin system also presents the same paradox as the particle/nucleus system: when and how would the system change from a state represented by a linear combination of |+x, -x> and |-x, +x> to a different state represented by only one of these or the other? (As we have seen, such paradoxes exist if state vectors have a one-to-one correspondence with states, but fail to exist if each of a multiplicity of concurrent state vectors represents a different aspect of any single state, the vectors being thus in a many-to-one correspondence with the states.)

We will take advantage of this mathematical similarity in the following to show how the interpretation of state vectors' significance presented here resolves further paradoxes illustrated by another two-constituent system.

## 6. EXPERIMENT OF SCULLY, ENGLERT, AND WALTHER

The well-known double-slit experiment, in which a single particle is associated with a wave of definite frequency that propagates through a pair of slits in a screen and produces an interference pattern, involves a phenomenon analogous in many respects to a single particle with spin. In this section, we take advantage of the analogy and apply some of our previous analysis to the double-slit experiment and some of its variations. In so doing we show how assumption of the many-to-one relationship between state vectors and physical states proposed in the preceding sections may resolve paradoxical features of these experiments, just as it appears to resolve paradoxes associated with the experiments discussed earlier.

To make the analogies more evident, let us again use the same kind of notation as was used for the single particle with spin. For the basic double-slit experiment, instead of having |+x> and |-x> describe spin directions parallel and antiparallel to the x-axis, let us use them to describe passage of a particle through the rightmost and leftmost slits respectively of the double-slit screen. Then we may represent the interference pattern produced by a wave whose phase is identical at the two slits by (|+x> + |-x>)/√2. (This type of wave is the one usually considered first; if the phases were different by π radians, the interference pattern would be represented by (|+x> - |-x>)/√2.) While the state vectors are analogous to those for spin-orientations of a single particle, the traditionally contemplated detector is dissimilar. What is determined is a position for the particle at a particular time. If one finds a particle at either slit at a particular instant, one directly determines which of the state vectors |+x> and |-x> properly characterizes it. But the way usually chosen[4,(10)] to determine that (|+x> + |-x>)/√2 (rather than (|+x> - |-x>)/√2) characterizes the wave is to determine the positions of many particles so characterized as they reach a detector some distance from the slits. The exact alternating spatial variation of particle-detection rates reveals which one wave-interference pattern is associated with all the particles.

By analogy with the case of a single particle with spin, we may symbolize the (|+x> + |-x>)/√2 interference pattern that characterizes every particle that passes through the system by "|+z>", and the orthogonal wave whose amplitude at the right slit is π radians out of phase with the amplitude at the left slit by "|-z>". It is impossible to

determine whether a particle's x-orientation is positive or negative while one is confirming that the particle has a positive-z spin, since a Stern-Gerlach apparatus can only be oriented along one direction at a time; it is just as impossible to determine whether |+x⟩ or |-x⟩ characterizes a particle passing through a double-slit screen if its interference pattern is "+z", but why this is impossible is less obvious.  The same Stern-Gerlach device can be used to measure a spin's orientation along any axis, but certain devices and means usually thought of for determining which slit a particle comes through are different from those that determine which interference pattern characterizes the wave.

In fact, it is possible, by using two separate detectors, to in some sense determine both which slit each particle passes through and which wave interference pattern characterizes it.[5,(11)]  But this means that the system is different; with two detectors to determine the two different features of the system, the system necessarily involves not one, but two constituents--a particle, and the detector used to find which slit the particle travels through.  Such a system exhibits behavior analogous to that of the two-constituent systems discussed previously.

One experiment of this type described by Scully, Englert, and Walther in *Nature*[(5,6)] involves projecting atoms one at a time toward a double slit, exposing them to laser photons *en route* that they can absorb and deflecting those atoms out of the system that don't absorb a photon, and having each remaining excited atom pass through a pair of maser cavities just before reaching the double slit, into one of which the atom spontaneously emits a microwave photon before going on.  The motion an atom has after absorbing the laser photon is not significantly disturbed when it emits its microwave photon in the cavity.  The electromagnetic field in each cavity is initially in the lowest energy state, so if the field inside one of them is found in a higher-energy oscillation mode, it is necessarily due to an atom that has passed through.  Since each cavity is in line with a different slit, one may check the path that any atom has taken by seeing in which cavity the electromagnetic field mode has changed. No interference pattern is definitely correlated with passage of an atom through either slit, so whether a change is found in the right or the left cavity, the atom that made it might be found anywhere beyond the slit screen; there are no forbidden "dark interference bands". But the cavities have a sensor between them which can be exposed to both cavities' interiors at the same time. If this sensor is so exposed, it will in due course either register the presence of a photon or fail to.

The intercavity photon detector is most likely to register a photon if the electromagnetic field between the two cavities is large, which implies that the electric (magnetic) fields are more or less parallel in the two cavities. If the electric (magnetic) field in one cavity is oppositely directed to that in the other, the fields cancel out between the cavities and the detector is far less likely to respond. As will be argued below, registration of a photon almost always correlates with an interference pattern for the atoms in which the amplitudes at both slits are in phase, and failure of the sensor to register a photon almost always correlates with the atom-interference pattern for which these amplitudes are $\pi$ radians out of phase. Thus by exposing the intercavity photon detector, we erase the evidence that the cavity modes provided of which slit the atom went through, but we gain the ability to tell almost certainly which interference pattern the atom would be found in--the one with both slits in phase, or the one with each slit $\pi$ radians out of phase with the other--if we checked the interference pattern.

How can the electromagnetic field mode be changed in only one of the cavities or the other, while the electromagnetic fields have equal nonzero strengths and parallel or antiparallel directions in the pair of cavities? We assume here for the field what we have assumed for particle characteristics in the previous examples: that different state vectors correspond to different features of the same state. Just as an atom may be characterized by a definite location and a particular interference pattern, so may an electromagnetic field. The state vector that determines the probability that the intercavity photon detector registers a photon describes whether electromagnetic fields in the two cavities are parallel or antiparallel, and is analogous to the state vector that describes an atom's interference pattern, while the state vector describing in which one of two cavities the electromagnetic field has been excited is a direct analogue of the position state vector for the atom.

Again, we can analyze this two-component (atom and microwave photon) system as we did above, and consider the implications of a state's being represented by multiple state vectors. Let "+x" represent either that an atom passes through the slit on the right or that a photon is emitted in the cavity on the right, and let "-x" represent the same for the slit and cavity on the left. If we were to let atoms that had not absorbed photons at all through the system, and directed them so that they were all characterized by a state vector whose position-representation wavefunction's amplitudes were in phase at both slits, the one state vector those atoms would all have in common would be

$$c(|+x\rangle + |-x\rangle)/\sqrt{2}.$$

The fact that we do not express this as $(c|+x\rangle + c'|-x\rangle)/\sqrt{2}$, with $|c| = |c'| = 1$ but otherwise arbitrary, signifies that each atom has the same amplitude at each slit and so interferes in a particular way. Following the convention we have used to describe spin-orientation eigenvectors and their interrelations, we could equivalently use the state vector $|+z\rangle$ to describe every atom in such a system. (According to the interpretation of state vectors presented earlier in this paper, individual atoms' other state vectors for other "directions of spin-orientation" will vary.)

Since we are dealing with excited atoms that emit one photon apiece in one of the two cavities, but whose motion is otherwise similar, their common state vector is

$$c(|+x, +x\rangle + |-x, -x\rangle)/\sqrt{2},$$

the first symbol in each ket corresponding to the atom and the second to which cavity's electromagnetic field is changed. This is partly demonstrable by the following consideration. If we determine which slit an atom passes through both by determining which cavity it leaves a photon in and by looking for the atom itself just past the slit, we will find the atom and the electromagnetic field excitation at the same slit--the pair of them characterized by either $|+x, +x\rangle$ or $|-x, -x\rangle$. Either of these occurrences is equally likely, and each has a 50% probability for a pair also characterizable by the state vector $c(|+x, +x\rangle + |-x, -x\rangle)/\sqrt{2}$; these probabilities would be different if the state vector common to all atom-field pairs included nonzero terms proportional to $|+x, -x\rangle$ or $|-x, +x\rangle$.

All that the foregoing demonstrates by itself is that the common state vector is $(c|+x, +x\rangle + c'|-x, -x\rangle)/\sqrt{2}$, with $|c|^2 = |c'|^2 = 1$. That c equals c' is demonstrable only by further consideration. Now if the unexcited-atom interference-pattern state vector $c(|+x\rangle + |-x\rangle)/\sqrt{2}$ were represented by a spin-direction, it would be the positive z-direction by our usual conventions. Let us use "+z" to stand for this interference pattern (amplitudes at both slits in phase), and "-z" to stand for the orthogonal pattern (amplitudes π radians out of phase). With equal fitness, "+z" and "-z" may respectively represent the "constructive" and "destructive interference" of electromagnetic-field wavefunctions whose amplitudes were in or out of phase in the two cavities. Thus $|+z, +z\rangle$ represents any situation in which both atom and electromagnetic-field wavefunction amplitudes are in phase; $|-z, -z\rangle$ represents the case of π-radian-out-of-

phase amplitudes for both atom and electromagnetic field; |+z, -z> represents constructive interference for the atom and opposite-phase amplitudes for the field; and |-z, +z> represents opposite-phase atom amplitudes with same-phase field amplitudes. If all four of these were compatible with the one state vector that all the atom-and-field pairs have in common, so that the situations represented by each of them had nonzero probability (amplitude), the phase factors c and c' would be unequal. The fact that neither |+z, -z> nor |-z, +z> is possible--that is, that the atom and electromagnetic-field amplitudes must both have the same phase relationships at the two cavities or slits-- means that c and c' are the same.

Why are |+z, -z> and |-z, +z> impossible? Consider how the phenomena they represent would have to occur. As we just noted, |+z, -z> describes the amplitude for the atom as being the same at each slit, while the electromagnetic-field amplitude has opposite phases at the two slits. This would mean that if the photon were emitted at the left cavity, the electric and magnetic field vectors would have one pair of orientations there, while if the photon were emitted in the right cavity, each of these field vectors would have the opposite orientations *there*; yet at each cavity, the emerging atom would have the same complex amplitude. The atom would thus have to have the same final state, whichever cavity it emerged from, even though the oscillating electromagnetic field in one cavity would have been in phase with the atom's oscillating charge while in the other, the field would have been out of phase. This is impossible, since emitted waves are always in step with the oscillations of their emitter. By a similar argument, we determine that the system can never be represented by |-z, +z>; this state vector describes electromagnetic fields in the two cavities that, on emission, turn out to have the same phase even though the atom would have come out with charge oscillation in step with the field at one cavity and out of step at the other. The scalar products of

(c|+x, +x⟩ + c'|-x, -x⟩)/√2 with |+z, -z⟩ and |-z, +z⟩ can be zero only if c and c' are equal; the products of (c|+x, +x⟩ + c'|-x, -x⟩)/√2 with |+z, +z⟩ and |-z, -z⟩ have the same nonzero amplitude in that case.  As calculation shows, this also means that d(|+z, +z⟩ + |-z, -z⟩)/√2 characterizes the common state vector just as well as c(|+x, +x⟩ + |-x, -x⟩)/√2, being an equivalent expression of it.

Thus the state vector |+z, +z⟩ only characterizes half of the excited atoms that come in and leave a photon.  |-z, -z⟩ characterizes the other half.  That is, if we were to expose both cavities to the intercavity photon sensor and locate each atom once it had passed beyond the slit screen, each atom whose microwave photon was registered by the sensor would most likely be found well within the "in-phase" interference pattern, in one of its antinodes, while any atom whose photon didn't register would probably be found well within an antinode of the orthogonal "π-out-of-phase" interference pattern, and seldom where this pattern's interference was destructive and the other's constructive. These photon and atom interference patterns would always be found together if both were looked for; both |+z, -z⟩ and |-z, +z⟩ are incompatible with the state vector c(|+x, +x⟩ + |-x, -x⟩)/√2 that characterizes every atom-and-emitted-photon pair, as the preceding discussion of the relevant probability amplitudes showed.

If, however, we do not expose the cavities to the sensor between them and erase the evidence of which cavity a photon was left in, but locate the electromagnetic field excitation in one cavity or the other, we find that the "in-phase" and "π-out-of-phase" interference patterns for the atom are equally likely for either location of the emitted photon, with either photon location being as likely as the other.  The multiple-state-vector/single-state interpretation of this is that whenever the state vector c(|+x, +x⟩ + |-x, -x⟩)/√2 characterizes all the atom-electromagnetic field pairs in this system, equal proportions of these atom-field pairs are also characterized by |+z, +x⟩,

|+z, -x>, |-z, +x>, and |-z, -x>.  Also, among the pairs characterized by each one of these vectors, equal proportions of them are further characterized by each of the vectors |+x, +z>, |+x, -z>, |-x, +z>, and |-x, -z>.  According to more usual interpretations of state vectors, appropriate measurements of the atom and electromagnetic field of a pair somehow change the pair's state from one represented by c(|+x, +x> + |-x, -x>)/√2 to one represented by one of these other vectors--which one depending of course on whether path or interference pattern properties are measured for the atom (at least to the extent that an interference pattern is measurable for a single atom) and for the electromagnetic field, but also depending on pure chance, which determines whether the values actually found are the "positive" or "negative" ones.  One may determine either the path or the interference pattern of the particles sent through the system, and one may decide after the photon is emitted, even after the atom is detected, whether one wants to go ahead and determine the atom's path with the detector or check the atom's interference pattern instead.

A major point of the Scully-Englert-Walther "eraser" experiment was to show that loss of an interference pattern does not imply uncontrolled disturbance of the measured object by whatever measures its properties, but rather can result from the fact that the measured system becomes correlated with what measures its properties.[6,(5,12)] Whereas only one interference pattern ((|+x> + |-x>)/√2, or |+z>) is found for atoms that absorb and emit no photons, so that an atom's path cannot be determined if it interferes since the atom leaves no tracks before interfering, those atoms that do absorb laser photons become part of a two-constituent (atom-microwave field) system, in which both atom and emitted photon can exhibit "|+z>-like" and "|-z>-like" phenomena, even though the motion of the atom is practically undisturbed when it emits the photon.  The atomic interference pattern turns out to be destroyed by the atom's interaction with the electromagnetic field that is used to find the atom's position, but it is "lost" only in a particular way; an atom correlated to a photon in the manner described may be characterized by either of two interference patterns, and so may the electromagnetic field itself.  Neither of these interference patterns is correlated with traversal of either slit in the screen.  But which slit the atom traverses is correlated with where it leaves its photon, and whether or not the electromagnetic field interacts with the unshielded intercavity detector is correlated with which interference pattern the atom appears in.[7]

In the experiment of Scully, Englert, and Walther, and in Bell's version of the Einstein, Podolsky, and Rosen experiment, the system constituents are correlated, so that as each constituent is measured at a different place, observation of that measurement tells something about the behavior of the other constituent.  By usual interpretations of quantum theory, predictions about one constituent's behavior based on the other's behavior are only about the first constituent's *potential* behavior--what *would* occur if the first constituent were to interact with appropriate apparatus.  This is more than a statement about what would be observed; it suggests that certain behavior does not happen, that certain features do not exist for the first constituent, unless an appropriate interaction takes place.  According to the ideas presented here about state vectors' compatibility, states themselves being characterized by a multiplicity of state vectors each, whatever is measured about one constituent of a two-constituent system always tells something about the other constituent, whether or not that constituent undergoes an interaction that can reveal it.  In effect, if a state vector like $(|+x, +x\rangle + |-x, -x\rangle)/\sqrt{2}$ or $(|+x, -x\rangle - |-x, +x\rangle)/\sqrt{2}$ characterizes a two-constituent system, either constituent can be considered a copy of the distant one (a kind of negative copy in the case of Bell's EPR experiment).

# 7. SYSTEM OF THREE PARTICLES WITH SPIN: PROBLEMS

In the previous sections, by concentrating on systems having only one or two constituents, we showed how the idea that any physical state requires multiple state vectors to completely represent it may resolve paradoxes inherent in the view that a single state vector alone represents the entire reality of any given physical system. Certain things are more interesting for quantum systems made of more than two constituents; for these, there is a greater difference between some more usual interpretations of their behavior and the interpretation outlined above. To explain this difference, we now proceed to consider a three-constituent system, first described by N. David Mermin, who based it on a thought experiment of Daniel Greenberger, Michael Horne, and Anton Zeilinger.[8,(13-15)] In this section, we describe this system and show the basic set of problems it raises for the interpretation of quantum physics. In the following section, we show a way to resolve the problems such systems present using the interpretation discussed previously, and point out how what this interpretation implies for such systems differs logically (though not experimentally) from what more usual interpretations imply.

The system in question is a system of three spin one-half particles. Initially together, with their center of mass at rest, the particles separate and travel in different directions with the total rectilinear momentum of the system remaining zero. A Cartesian coordinate system is defined for each particle, such that each particle's z-axis is the direction of its momentum, +z being the direction the particle travels while -z is the opposite direction. For each particle, the x-axis is defined as the axis perpendicular to all three particles' momenta, such that all three x-axes are positive in the same direction. The y-axes' directions are thus coplanar with the particles' momenta; we choose their orientations so that each particle's coordinate system is right-handed. See Figure 1.

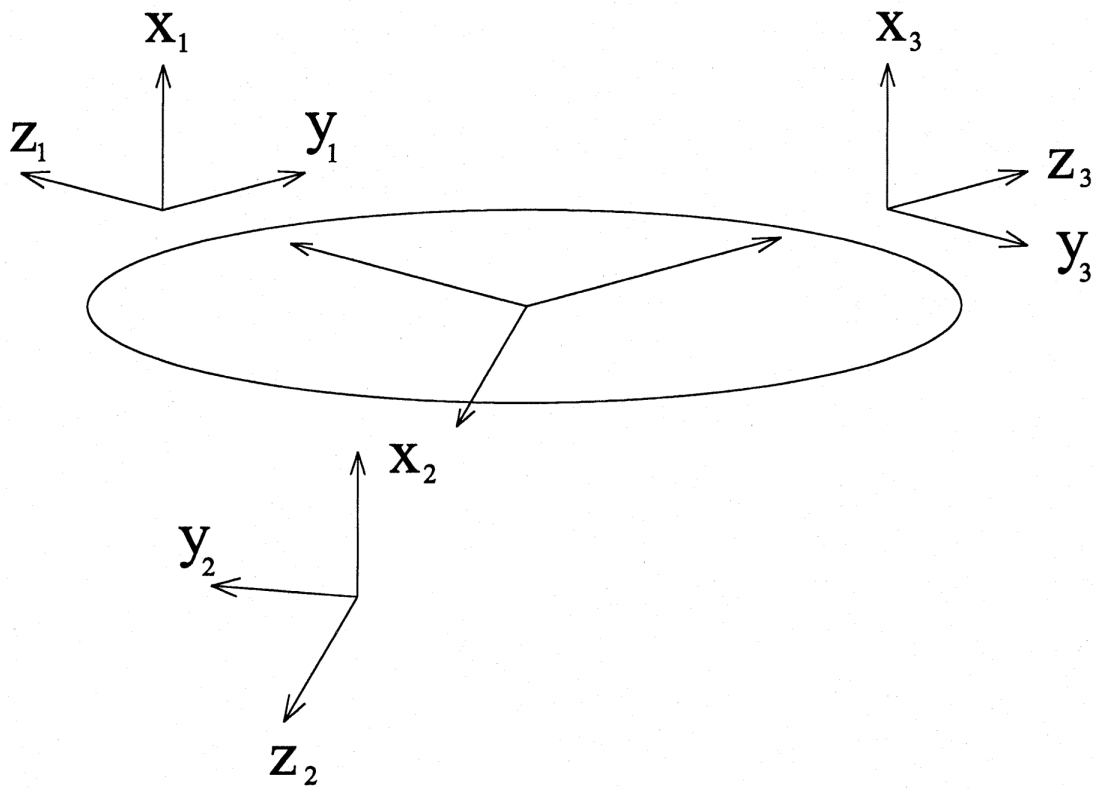

Figure 1. Momentum vectors of particles in a three-particle Greenberger-Horne-Zeilinger experiment and the corresponding coordinate systems, one for each particle.

The system is characterized by the state vector $(|+z, +z, +z\rangle - |-z, -z, -z\rangle)/\sqrt{2}$, in which the +zs and -zs in each term describe spin-orientations of the first, second, and third particles, respectively. This state vector is an eigenvector of four operators:

$$(\sigma_1 \cdot x)(\sigma_2 \cdot y)(\sigma_3 \cdot y),$$

$$(\sigma_1 \cdot y)(\sigma_2 \cdot x)(\sigma_3 \cdot y),$$

(16)

$$(\sigma_1 \cdot y)(\sigma_2 \cdot y)(\sigma_3 \cdot x),$$

$$(\sigma_1 \cdot x)(\sigma_2 \cdot x)(\sigma_3 \cdot x),$$

with the corresponding eigenvalues being +1, +1, +1, and -1.[9] Note that $(|+z, +z, +z\rangle - |-z, -z, -z\rangle)/\sqrt{2}$ is an eigenvector of these *products* of individual-particle spin operators; it is not an eigenvector of any single one of the individual-particle operators themselves.

For this thought experiment, we have one Stern-Gerlach apparatus stationed to intercept each of the three particles with which we measure the spin-direction along some axis perpendicular to the particle's momentum--i.e., somewhere in its x-y plane. Suppose the apparati are oriented to measure the spin-orientation along the x-axis for the first particle and along the y-axes for the second and third particles. All that is certain about the outcome is that the product of the spin-orientations--+1 if parallel to an axis and -1 if antiparallel--will be +1, and that the state vector found will be one of the four of type $|\pm x, \pm y, \pm y\rangle$ with the signs independent of each other except for the constraint on their combined product, and not of a type like $(|+z, +z, +z\rangle - |-z, -z, -z\rangle)/\sqrt{2}$, since what is found must be an eigenvector of the operators $(\sigma_1 \cdot x)$, $(\sigma_2 \cdot y)$, and $(\sigma_3 \cdot y)$ *individually*. By similar reasoning, if the Stern-Gerlach devices are oriented so that the second measures the x-component of spin while the others measure y-components, or the third measures the x-component while the first two measure y-components, or all three measure their particles' spins' x-components, we may see that the state vectors found will be of the type |±y, ±x, ±y>, |±y, ±y, ±x>, or |±x, ±x, ±x>, with signs constrained only so that the spin-orientations' products are +1, +1, or -1 respectively, according to how the Stern-Gerlach devices are oriented.

It is not possible to set up the same collection of Stern-Gerlach apparati to measure more than one of these sets of spin-orientations for the same three-particle system, but we could set them to any one orientation we like for observing any single system. Suppose the system were in a state such that, just as (|+z, +z, +z> - |-z, -z, -z>)/$\sqrt{2}$ characterized it of course, we would also find, for appropriate Stern-Gerlach device orientations, that the state vectors |-x, +y, -y>, |+y, -x, -y>, and |+y, +y, +x> also characterized it. These state vectors are eigenvectors of the respective spin-operator combinations $(\sigma_1 \cdot \mathbf{x})(\sigma_2 \cdot \mathbf{y})(\sigma_3 \cdot \mathbf{y})$, $(\sigma_1 \cdot \mathbf{y})(\sigma_2 \cdot \mathbf{x})(\sigma_3 \cdot \mathbf{y})$, and $(\sigma_1 \cdot \mathbf{y})(\sigma_2 \cdot \mathbf{y})(\sigma_3 \cdot \mathbf{x})$ as well as of their individual-factor operators; their eigenvalues with respect to the product operators are the same as those of the state vector (|+z, +z, +z> - |-z, -z, -z>)/$\sqrt{2}$ that describes all the three-particle systems of this experiment, and so are consistent with it--indeed, there are definite nonzero probabilities that these state vectors describe the same system; and the individual spin-orientations in the x- and y-directions are consistent among those vectors for which those particular orientations are definite, again shown by the fact that there are nonzero probabilities that both members of the different possible pairs among these three state vectors characterize the same system.

A problem exists at this point. For this example, we have not yet found a state vector of the form $|\pm x, \pm x, \pm x\rangle$ whose eigenvalue with respect to $(\sigma_1 \cdot \mathbf{x})(\sigma_2 \cdot \mathbf{x})(\sigma_3 \cdot \mathbf{x})$ is the same as that of $(|+z, +z, +z\rangle - |-z, -z, -z\rangle)/\sqrt{2}$. There are only four such possible eigenvectors of the individual $(\sigma_n \cdot \mathbf{x})$--$|+x, +x, -x\rangle$, $|+x, -x, +x\rangle$, $|-x, +x, +x\rangle$, and $|-x, -x, -x\rangle$--that have the same eigenvalue (-1) for the single operator $(\sigma_1 \cdot \mathbf{x})(\sigma_2 \cdot \mathbf{x})(\sigma_3 \cdot \mathbf{x})$. Three of these are compatible with two of the state vectors $|-x, +y, -y\rangle$, $|+y, -x, -y\rangle$, and $|+y, +y, +x\rangle$ but not with the third; the remaining one is compatible with none of them! The only "all x" state vector consistent with $|-x, +y, -y\rangle$, $|+y, -x, -y\rangle$, and $|+y, +y, +x\rangle$ is $|-x, -x, +x\rangle$, but its eigenvalue with respect to $(\sigma_1 \cdot \mathbf{x})(\sigma_2 \cdot \mathbf{x})(\sigma_3 \cdot \mathbf{x})$ is +1, and it is incompatible with (orthogonal to) the one state vector $(|+z, +z, +z\rangle - |-z, -z, -z\rangle)/\sqrt{2}$ that is common to all the three-particle systems in our experiment. Given the assumptions about compatibility among state vectors described above, there is thus no eigenvector of $(\sigma_1 \cdot \mathbf{x})(\sigma_2 \cdot \mathbf{x})(\sigma_3 \cdot \mathbf{x})$ of the "$|x, x, x\rangle$" type that can describe this system. Note that any of the "x" and "x-y" state vectors with appropriate eigenvalues is compatible with the "z" vector $(|+z, +z, +z\rangle - |-z, -z, -z\rangle)/\sqrt{2}$ alone; the problem is that there are no sets of state vectors that include one for every measurable combination of x- and y-orientations whose members are each consistent with all the others.

This problem, as one may work out, is not unique to the particular choice of which spin-orientations might be found in an experiment, or to which types of state vectors--"$|x, y, y\rangle$", "$|y, x, y\rangle$", "$|y, y, x\rangle$", or "$|x, x, x\rangle$"--are to be certain of inclusion in a consistent system. Whether it is a state vector of the "all x" form or one of the "two ys, one x" type, at least one of these types of vector will have four possible forms consistent with $(|+z, +z, +z\rangle - |-z, -z, -z\rangle)/\sqrt{2}$ but inconsistent with at least one among the vectors of the other three types that might characterize the system, no matter what these other three vectors themselves might be. As we have seen, if a consistent set of "|x, y, y⟩", "|y, x, y⟩", and "|y, y, x⟩" vectors is chosen, no "|x, x, x⟩" vector consistent with all of them will also be consistent with (|+z, +z, +z⟩ - |-z, -z, -z⟩)/√2; but had the system instead been characterized by, say, |+x, +x, -x⟩, |+y, +x, +y⟩, and |+y, -y, -x⟩ instead, we would have been equally stuck with no possible "|x, y, y⟩" vector consistent with all the features of the system's state.

The similar discussion presented in Mermin's paper indicates that *local* hidden variables such as Bell discussed, which in combination with the state vector would determine the x- and y-orientations of each particle's spin, would result in phenomena inconsistent with quantum theory; whatever values are chosen for all these orientations, at least one combination of them is incompatible with a known feature of the state of the system. This is no problem for quantum theory in a more usual interpretation; some eigenvector is possible for whatever combination of x- and y-spin-orientations one might look for, and one can never see more than one such combination at a time anyway, so according to that interpretation, the question of what *un*observed combinations might or might not be compatible with what is observed is of no practical significance. We might expect that something will be observed for whatever orientation of Stern-Gerlach apparati one sets up. But then one still has the problem of how a particular eigenvector of the observing apparatus comes to be the real one, with the attendant paradoxes considered at the beginning of this paper.

On the other hand, we see that even if a multiplicity of state vectors characterizes any one physical state, then the reasonable expectation, that state vectors corresponding to a *definite* result for the measurement of any specific quantity (such as the spin-orientation of one particle along a particular axis) all correspond to the *same* result if all those vectors together describe one state, presents a problem. Is there any way to resolve it?

To see further what the problem does and does not consist of, let us consider again the set of state vectors (|+z, +z, +z⟩ - |-z, -z, -z⟩)/√2, |-x, +y, -y⟩, |+y, -x, -y⟩, and |+y, +y, +x⟩. These four are consistent with each other--the probability that any scalar product of them ⟨Φ|Ψ⟩ characterizes any one of the three-spin systems under consideration is nonzero. But none of the eight eigenvectors of the individual x-orientations of all three spins is compatible with all four. Suppose, then, that no possible measurement of all three x-spin-orientations exists--that if we tried to measure the x-direction of all three particles' spins, we would not find them all--since such a measurement would correspond to one of the impossible state vectors. Could the four state vectors (|+z, +z, +z⟩ - |-z, -z, -z⟩)/√2, |-x, +y, -y⟩, |+y, -x, -y⟩, and |+y, +y, +x⟩ together describe compatible aspects of a single state? By the aforementioned consistency assumptions, no. The four state vectors are indeed compatible with each other; if any one of them describes the physical system, each of the others individually is at least a possibility, so far as the one state vector alone is concerned. Each of the latter three describes what could happen if two spins' y-orientations and one spin's x-orientation were measured, and each is consistent with what the first state vector implies about the spin-orientations' products. But suppose we tried to measure all three spins' x-orientations in this case. Each of the latter three vectors implies a specific outcome for each of these spin measurements. If no information is transmitted among Stern-Gerlach apparati about each others' orientations, the result of measuring a spin's x-orientation at one apparatus should be the same whether the other apparati are set to measure x-orientations, y-orientations, or any other orientations. Thus if all three state vectors |-x, +y, -y⟩, |+y, -x, -y⟩, and |+y, +y, +x⟩ correctly described inevitable outcomes of appropriate measurements, the result of turning all three Stern-Gerlach devices to measure the three spins' x-orientations would be to show that for particles 1 and 2 these were negative, while for particle 3 it was positive--just as surely as if |-x, -x, +x⟩ were a

"real" state vector, and about as surely impossible since the system is also characterized by $(|+z, +z, +z\rangle - |-z, -z, -z\rangle)/\sqrt{2}$.

Then suppose instead that one of the vectors $|-x, +y, -y\rangle$, $|+y, -x, -y\rangle$, or $|+y, +y, +x\rangle$ did not characterize the state of the system either. For the purpose of discussion, it does not matter which is omitted, but for definiteness assume that the first is not a feature of the state. That leaves $|+y, -x, -y\rangle$ and $|+y, +y, +x\rangle$ to characterize the system along with $(|+z, +z, +z\rangle - |-z, -z, -z\rangle)/\sqrt{2}$ (together, of course, with state vectors corresponding to other spin-orientations that could be measured but are not discussed in detail here).

Now consider what these three state vectors together imply about what we would find were we to try to measure the x-orientation of the first particle's spin. If we measured the x-orientations of spins 2 and 3 at the same time, we would expect that, since these latter results would be -x and +x respectively, x-orientation 1 would be +x so the product of all three would come out right. But if we measured y-orientations 2 and 3 instead, we would get +y and -y, and therefore expect to see -x for the first particle's x-spin-orientation. However, no such problem would exist if we saw neither x-orientation for the first particle.

Contradictions would not exist if the set of state vectors that described real features of this system's state did not include any state vectors for values of at least one of the particles' spin-orientations. Here we have considered an absence of state vectors that directly implied either positive or negative x-orientations for particle 1's spin, but other possible states of this system might involve such absence for a different particle and/or an orthogonal spin-axis, so that the set of state vectors that together describe the state lacks any vectors that imply specific values therefor. For example, some different states of this system might be described by sets of state vectors that included some for all combinations of x and y spin-orientations except for the y-axis of the second particle. For such states we would expect to find that orienting the second Stern-Gerlach apparatus along the second y-axis resulted in no reading whatsoever.

At first sight, this may appear to imply a very significant departure from the assumptions of quantum physics. It seems generally taken for granted that whenever a quantum-physical system is measured to determine some quantity, some value will be found for it. If the "original" or the originally known state vector implied no definite value for some measurement, the system would turn out to manifest one anyway, perhaps through "collapse" of the system's state vector or wave function. Here we assume that in some cases, a measurement results in no finding whatsoever. But how significant is this?

First, in the system just discussed, not finding a spin-orientation may seem to mean that a particle is somehow missing, with perhaps serious implications for momentum conservation in the system studied. But such an occurrence needn't mean that no particle passed through the Stern-Gerlach apparatus, only that it didn't register as having a definite spin along the axis examined. Had we measured the particle's momentum instead, in this circumstance we might well have found it. Second, we see that of the four forms of "individual x- and y-orientation state vectors" under consideration, only two of them would have to not exist. This suggests that only in some cases would we fail to measure a value for something we might try to observe. The question is: How much would we not be able to measure?

## 8. SYSTEM OF THREE PARTICLES WITH SPIN:  RESOLUTION OF PROBLEMS

From the foregoing, it may seem that one would fail to measure definite values for particular variables fairly often.  But we have only considered spin-orientations along two different axes for each particle.  There are many other axes within the x-y plane, and many others still not even in the x-y plane.  This fact turns out to be quite significant.  Our problem is that sets of state vectors, consistent with one another two at a time, are inconsistent all taken together.  In this section, we resolve the problem and show how the resolution's implications differ logically from those implied by more usual interpretations of quantum physics.

There would be no problem if it weren't for the fact that the system in question is in a state properly characterized by $(|+z, +z, +z\rangle - |-z, -z, -z\rangle)/\sqrt{2}$.  Self-consistent combinations that include one state vector each of the "$|x, y, y\rangle$", "$|y, x, y\rangle$", "$|y, y, x\rangle$", and "$|x, x, x\rangle$" types are easily put together.  But at most only two vectors from each such set will be consistent with each other *and* with $(|+z, +z, +z\rangle - |-z, -z, -z\rangle)/\sqrt{2}$, assuming that all state vectors for the actual system that imply a definite value for any single measurement all imply the same value for that measurement.  Thus we may begin to determine which state vectors are impossible by working out which eigenvectors of individual spin-orientation operators are completely incompatible with $(|+z, +z, +z\rangle - |-z, -z, -z\rangle)/\sqrt{2}$.

Consider all directions in three-dimensional space, not just those in the x-y coordinate planes of each of the three particles.  Let $|n_1, n_2, n_3\rangle$ represent a state vector describing what we would find if all three particles had positive spins in the set of directions $\mathbf{n}_1$, $\mathbf{n}_2$, and $\mathbf{n}_3$ and we measured them.  We want to know when the amplitude $\langle n_1, n_2, n_3|(|+z, +z, +z\rangle - |-z, -z, -z\rangle)/\sqrt{2}$ would be equal to zero. If we represent the operator for spin-orientation along the *i*th n-axis as $\sigma \cdot \mathbf{n}_i$ and specify the direction of **n** by the usual spherical coordinates $\theta$ and $\phi$, we have

$$\sigma \cdot \mathbf{n}_i = \begin{bmatrix} +\cos\theta_i & +\sin\theta_i \cdot e^{-i\phi_i} \\ +\sin\theta_i \cdot e^{+i\phi_i} & -\cos\theta_i \end{bmatrix}. \tag{17}$$

The eigenspinor for this whose eigenvalue is +1 is (up to an arbitrary phase factor)[10]

$$\frac{1}{\sqrt{2}} \begin{bmatrix} \sin\theta_i / \sqrt{1-\cos\theta_i} \\ e^{+i\phi_i} \cdot \sqrt{1-\cos\theta_i} \end{bmatrix}. \tag{18}$$

Taking appropriate limits as $\theta \to 0$ and $\theta \to \pi$, we find that the eigenvectors of $\sigma \cdot \mathbf{z}_i$ are $[1\ 0]^T$ (for eigenvalue +1) and $[0\ 1]^T$ (for eigenvalue -1). Representing $|+z, +z, +z\rangle$ as $|[1\ 0]^T, [1\ 0]^T, [1\ 0]^T\rangle$, $|-z, -z, -z\rangle$ as $|[0\ 1]^T, [0\ 1]^T, [0\ 1]^T\rangle$, and $|n_1, n_2, n_3\rangle$ in general as

$$\frac{1}{\sqrt{2^3}} \left| \begin{bmatrix} \sin\theta_1 \sqrt{1-\cos\theta_1} \\ e^{+i\phi_1} \cdot \sqrt{1-\cos\theta_1} \end{bmatrix}, \begin{bmatrix} \sin\theta_2/\sqrt{1-\cos\theta_2} \\ e^{+i\phi_2} \cdot \sqrt{1-\cos\theta_2} \end{bmatrix}, \begin{bmatrix} \sin\theta_3/\sqrt{1-\cos\theta_3} \\ e^{+i\phi_3} \cdot \sqrt{1-\cos\theta_3} \end{bmatrix} \right\rangle. \tag{19}$$

we find that if

$$\langle n_1, n_2, n_3|(|+z, +z, +z\rangle - |-z, -z, -z\rangle)/\sqrt{2} = 0, \tag{20}$$

then

$$\frac{1}{4}\begin{pmatrix} \dfrac{\sin\theta_1 \cdot \sin\theta_2 \cdot \sin\theta_3}{\sqrt{(1-\cos\theta_1)(1-\cos\theta_2)(1-\cos\theta_3)}} \\ \\ -e^{-i(\phi_1+\phi_2+\phi_3)} \cdot \sqrt{(1-\cos\theta_1)(1-\cos\theta_2)(1-\cos\theta_3)} \end{pmatrix} = 0, \quad (21)$$

or

$$\prod_{i=1}^{3}\left(e^{+i\phi_i} \cdot \frac{\sin\theta_i}{1-\cos\theta_i}\right) = 1. \quad (22)$$

Examining this last equation shows us that a state vector $|n_1, n_2, n_3\rangle$ is incompatible with $(|+z, +z, +z\rangle - |-z, -z, -z\rangle)/\sqrt{2}$, or has zero probability of characterizing any system characterized by Examining this last equation shows us that a state vector $|n_1, n_2, n_3\rangle$ is incompatible with $(|+z, +z, +z\rangle - |-z, -z, -z\rangle)/\sqrt{2}$, or has zero probability of characterizing any system characterized by $(|+z, +z, +z\rangle - |-z, -z, -z\rangle)/\sqrt{2}$, only if $\prod_{i=1}^{3} e^{+i\phi_i} = 1$ (that is, if $\phi_1, \phi_2,$ and $\phi_3$ add together so that the phase of $\prod_{i=1}^{3}\left[e^{+i\phi_i} \cdot \sin\theta_i/(1-\cos\theta_i)\right]$ is zero) and $\prod_{i=1}^{3}\left[\sin\theta_i/(1-\cos\theta_i)\right] = 1$ (so that the magnitude is unity).

Suppose that our three-spin system, characterized by $(|+z, +z, +z\rangle - |-z, -z, -z\rangle)/\sqrt{2}$, were also characterized by eigenvectors of the individual operators $\sigma_1 \cdot n_1$, $\sigma_2 \cdot n_2$, and $\sigma_3 \cdot n_3$ for all combinations of $n_1$, $n_2$, and $n_3$ *except* for the $n_i$ that lay in their respective x-y planes--that is, for any orientation of a Stern-Gerlach apparatus *other* than one in the x-y plane, one would definitely find a value of $\sigma_i \cdot n_i$. If this were so, the problem with some of the state vectors for x-y plane spin-orientations would be insignificant. For in practice, the chances of aligning any one of the Stern-Gerlach apparati exactly in the x-y plane of its particle are zero. There is a nonzero probability of alignment within some noninfinitesimal neighborhood of this plane, but an infinitesimal probability of alignment exactly within it. Thus, if *only* state vectors that specified spins in the particles' x-y planes were lacking, their absence would never be noticed. As a practical matter, one would never check for their presence anyway; one might always check spin-orientations *arbitrarily near* x-y planes, or for that matter arbitrarily near any specific orientation. It is certain that one can examine spin-orientations along *some* direction in space; but though it is likely that one can check them near any *particular* axis, it is impossible to look for them precisely along any prespecified one.[11,(16)]

If self-consistent physical states could be described by sets of state vectors that failed to specify spin-orientations only in such an infinitesimally few directions, we would thus observe nothing different from what we would expect according to our more usual views of quantum physics. But we would have an explanation for the three puzzles mentioned at the beginning of this discussion. If the precise state of whatever system we were observing were specified by particular state vectors for anything we might observe, there would be nothing capricious, like a random collapse of the first state vector we had observed (or a corresponding wavefunction), to result in whatever else we observed in the system that was not defined by that first state vector; the result would simply correspond to one of those other state vectors that did define the outcomes of such observations. With no collapse, there would be no question as to where and when, in the chain of physical processes between the phenomenon and our observation of it, the first state vector collapsed into an eigenvector of the observed quantity's operator. And the fact that all observations throughout space fit the newly observed state vector at the same time would not imply that somehow our own frame of reference, which determines for us what events are simultaneous throughout all space, was special, if one state vector existed all along for every possible combination of variables we might observe, so that the set of all those state vectors combined, not just one of them (e.g., the first or last one we observe), together determined the system's state.

Could a self-consistent set of state vectors exist that specified all but an unobservably, infinitesimally few variables? In this case, at least, one set could--in fact, several could, corresponding to different states of a three-spin system that is characterized in part by (|+z, +z, +z⟩ - |-z, -z, -z⟩)/√2.

Two such states are easily described. Let the first particle's spin-orientation be "toward the first particle's positive z-direction" for every axis that is not in the first particle's x-y plane; that is, let the eigenvalue of any eigenvector of $\sigma_1 \cdot \mathbf{n}_1$ characterizing the system be +1 if $\mathbf{n}_1 \cdot \mathbf{z}_1 > 0$ and -1 if $\mathbf{n}_1 \cdot \mathbf{z}_1 < 0$. Let the same be true for the second and third particles as well. Then if the particles' spin-orientations are checked along any axes not lying in the x-y plane, they will always be in directions such that $\theta_1$, $\theta_2$, and $\theta_3$ are all less than $\pi/2$. Under these circumstances, the product $\prod_{i=1}^{3}\left[\sin\theta_i/(1-\cos\theta_i)\right]$ could never equal unity, but would always be greater. Thus a collection of state vectors, among which all that were eigenvectors of any given $\sigma_i \cdot \mathbf{n}_i$ all had the same eigenvalue for it, could be entirely self-consistent, provided that this collection did not include state vectors for which $\mathbf{n}_i$ was in the ith particle's x-y plane. The other easily described state is just like the first, except that the eigenvalues of the $\sigma_i \cdot \mathbf{n}_i$ are all -1 for $\sigma_i \cdot \mathbf{n}_i > 0$ and +1 for $\sigma_i \cdot \mathbf{n}_i < 0$--that is, each observable spin-orientation would be found "toward its particle's negative z-direction" if measured. $\prod_{i=1}^{3}\left[\sin\theta_i/(1-\cos\theta_i)\right]$ in this case would always be less than unity.

If these were the only two states for three-spin systems characterized by (|+z, +z, +z⟩ - |-z, -z, -z⟩)/√2, we would never see spin combinations for which one particle had spin-orientation θ in one "hemisphere" while the other two particles' spins

were aligned toward the other "hemisphere". But ordinary quantum-theory calculations show us that the probability for such an occurrence is not zero (unless the θs are 0 and π, with one spin being in one of these directions and the other two in the other). So if self-consistent sets of state vectors are to represent physical states, there should be some self-consistent sets that include state vectors of this sort. And there are; indeed, there are an infinite number of such sets in which any "three-spin" state vector, other than one for which $\prod_{i=1}^{3}\left[\sin\theta_i/(1-\cos\theta_i)\right]=1$, may be included, and such sets can be easily constructed.

Pick two of the three particles, and choose an angle $\theta_i$ for one and an angle $\theta_j$ for the other, both between 0 and π (it being indifferent whether $\theta_i$ does or does not equal $\theta_j$). For each of the two particles, consider the following four solid-angle regions: (1) the one between π/2 and $\theta_i$ ($\theta_j$); (2) the region "antipodal" to the first one, between π/2 and π - $\theta_i$ (π - $\theta_j$); (3) the region between the angle $\theta_i$ ($\theta_j$) and the "pole" 0 or π, whichever one is nearest; and (4) the region between π - $\theta_i$ (π - $\theta_j$) and the remaining pole. For the remaining particle, there will be an angle $\theta_k$ such that the product of the three "sin θ/(1 - cos θ)"s for $\theta_i$, $\theta_j$, and $\theta_k$ equals unity. Consider four regions similarly defined by $\theta_k$ for the remaining particle as well.

A self-consistent set of state vectors of the type |$n_1$, $n_2$, $n_3$> can be found as follows. For each particle, choose one pole (either θ = 0 or θ = π). Let all orientations between that pole and the nearest boundary ($\theta_i$, $\theta_j$, $\theta_k$, π - $\theta_i$, π - $\theta_j$, or π - $\theta_k$, as the case may be) be possible spin-orientations for that particle. The next region, from this boundary to the "equator" π/2, will encompass spin-orientations that will *not* be possible for the particle. The next solid-angle region, encompassing those orientations whose latitudes lie between π/2 and the next boundary (π - $\theta_{(i,j,k)}$ or $\theta_{(i,j,k)}$), represents the remaining possible spin-orientations for that particle, while the other region (between the

boundary $\pi - \theta_{(i,j,k)}$ or $\theta_{(i,j,k)}$ and the other pole) includes other directions in which the spin-orientation for that particle will not be found. All state vectors, which are eigenfunctions of all three operators $\sigma_1 \cdot \mathbf{n}_1$, $\sigma_2 \cdot \mathbf{n}_2$, and $\sigma_3 \cdot \mathbf{n}_3$ for all $\mathbf{n}_1$, $\mathbf{n}_2$, and $\mathbf{n}_3$ having directions in these regions (thus excluding, by construction, any triplets $[\mathbf{n}_1, \mathbf{n}_2, \mathbf{n}_3]$ such that $\prod_{i=1}^{3}\left[\sin\theta_i/(1-\cos\theta_i)\right] = 1$), and are consistent with the spin-orientations so chosen, combined with either $|+z, +z, +z\rangle$ or $|-z, -z, -z\rangle$, together comprise a self-consistent set of state vectors. Spin-orientations are thus undefined only for the infinitesimal region of solid angles along the boundary directions $\theta_i$, $\pi - \theta_i$, and $\pi/2$ for each particle. See Figure 2. (If $\theta_i$ (and/or $\theta_j$ and/or $\theta_k$) $= \pi/2$, we have the even simpler situation of only two solid-angle regions to consider for that/those particles instead of four, with a "smaller" infinitesimal set of undefined spin-orientations.)

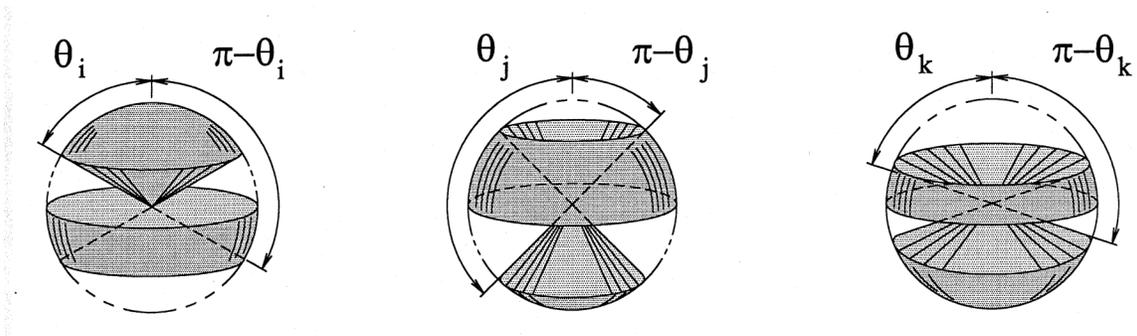

Figure 2. One set of possible spin-orientations for a self-consistent set of state vectors, corresponding to particular choices for $\theta_i$ and $\theta_j$ (with consequent value for $\theta_k$). Here, the ranges $0 < \theta < \theta_i$ and $\pi/2 < \theta < \pi-\theta_i$ include possible spin-orientations for particle i, angles $\theta$ such that $\pi-\theta_j < \theta < \pi/2$ and $\theta_j < \theta < \pi$ are possible spin-orientations for particle j, and the ranges $\pi-\theta_k < \theta < \pi/2$ and $\theta_k < \theta < \pi$ contain possible values of $\theta$ for particle k's spin-orientation. Eigenvectors of the individual operators $\sigma_i \cdot \mathbf{n}_i$, $\sigma_j \cdot \mathbf{n}_j$, and $\sigma_k \cdot \mathbf{n}_k$ for unit vectors $\mathbf{n}$ whose angles $\theta$ with their particles' respective z-axes lie within these ranges are consistent with each other and with $(|+z, +z, +z\rangle - |-z, -z, -z\rangle)/\sqrt{2}$.

Thus we see that even for systems of three components, each pair of which can be said to "measure" the other component (just as in our two-component systems either component could serve as a means of "depicting" the other component), it is possible to resolve the puzzles arising from our familiar conceptions of quantum behavior. The concept of compatible, coexisting state vectors that together comprise a system's state has a logical consequence for three-spin systems that differs from more familiar interpretations, but this logical consequence is not a practical, observational consequence. In either sort of interpretation, one can observe, infinitesimally often, state vectors infinitesimally far from the aforementioned "boundaries" that are very nearly opposite to any of the state vectors that are incompatible with the "initial" or "initially known" state vector. This has no observational consequences that are practically different from those of a different assumption--that some state vector exists (or represents something one would find with the appropriate observation) in a "boundary" region, even though it would be incompatible with the "initial" vector according to standard quantum-theory assumptions about probabilities. While practically no different from the ideas presented here, however, such an assumption is logically more complicated, in that for just these cases we would assume that orthogonal state vectors may be part of a single set that completely describes a physical state.

## 9. EXPERIMENT OF ZOU, WANG, AND MANDEL

The concepts used to address the paradoxes of the systems considered so far do have more practically testable consequences. Another experiment involving two related components, which at first sight is quite similar to that of Scully, Englert, and Walther, was performed by Zou, Wang, and Mandel at the University of Rochester and reported in *Physical Review Letters*.(17) Its behavior, however, presents a puzzle not present in the Scully-Englert-Walther experiment. In the latter system, the behavior of the atoms is independent of whether one sets up the apparatus to determine where the microwave photon was emitted or to determine whether the electromagnetic field has the same or opposite phases in the two cavities. Regardless of what aspect of the field's state the system is set to reveal, there is no single interference pattern common to all the atoms passing through the double slit. But the Zou-Wang-Mandel experiment is different. Like the Scully-Englert-Walther experiment, it involves a single entity that is converted into two, each of which can be examined to determine either its source or its "relative phase distribution" between source locations. But in the Zou-Wang-Mandel experiment, one of the two entities may exhibit an interference pattern--the visibility of which depends on how well the other entity's source can be determined given the particular way the apparatus is set up. If where the second entity came from is indeterminable, a unique interference pattern is found for the first entity. If the apparatus is altered so that the second entity's origin can be precisely determined, the first entity has no unique interference pattern, just like the atoms in the Scully-Englert-Walther experiment.

It is possible to derive a correct formula for these phenomena from a set of assumptions that themselves do not contradict usual interpretations of quantum physics in an obvious way, but that do present puzzles of an "action-at-a-distance" variety, and also appear problematic in terms of the interpretation presented in this paper. In this section we will derive the formula from a different basis consistent with the interpretation presented above, and show testable consequences of the new assumptions.

To appreciate the difficulty, let us see how the Zou-Wang-Mandel apparatus works. This system splits a laser beam two ways, and sends each of the resulting beams through a down-converter, which converts each incident laser photon into two photons that share the original photon's energy, thus producing two beams from each down-converter. One beam from each down-converter (the "signal") is directed so that both converge from different directions onto the same single detector. The remaining beams are aimed along the same direction, so that the leftover (or "idler") beam from the first down-converter, if not blocked, goes into the path of the idler from the second converter, both beams then striking another single detector in that direction. When all four beams from the two down-converters proceed to their targets unhindered by other objects, the two signal beams, which have the same frequency, produce an interference pattern, which is the same for all signal photons. But if the first idler beam is blocked before reaching the second down-converter, this interference pattern disappears.

Such an effect would seem reasonable if the first idler beam stimulated down-conversion in the second converter, thereby making it produce signal photons that were coherent with those from the first down-converter. But the effect exists even when the intensity of the original laser beam is so low that only one signal photon and one idler photon are expected to be in the apparatus at any given time. Under such conditions, no idler photon produced in the first down-converter can induce down-conversions in the second; induced down-conversion would result in two idler photons and two signal photons, not one of each.

It would thus seem that the signal beams are instantaneously affected by some remote, distant action on the path of one of the idler beams. It is as though, in the Scully-Englert-Walther experiment, separating the microwave cavities to allow determination of the paths taken by individual atoms, or relinking the cavities (with or without the intercavity photon detector), somehow made a difference in what the atoms themselves did beyond the cavities and the double slit--which, if indeed there is no influence of the first idler beam on the second down-converter, would bring us back to the third puzzle of quantum theory cited in the introduction: apparent action at a distance, instantaneous in only one reference frame that is otherwise undistinguishable from all other frames. (This form of the puzzle is a bit different from that discussed earlier, in that the apparent instantaneous change in the system relates to changing the sort of observation one can perform rather than the actual making of the observation.)

Zou, Wang, and Mandel derive a formula that accurately describes the dependence of the interference pattern's visibility on the transparency of the idler-beam path between the first and second down-converters, but the derivation is based on an assumption that in effect means the first down-converter may produce an idler photon but no signal photon, while the second down-converter produces a signal photon but no idler photon.[12,(18)] This down-conversion of a spatially-distributed incident photon into one idler and one signal of decidedly local and separate origins would seem to be about as paradoxical as the effect to be explained by it, being still another kind of instantaneous action-at-a-distance phenomenon lacking obvious consistency with special relativity (as well as with local energy conservation), presenting us with yet another variation of our third puzzle. Paradoxical or not, one could check for such a phenomenon by experiment: simply see whether an idler photon were ever produced with no signal photon at the first down-converter while a signal photon with no idler were produced at the second down-converter.

But regardless of how the observed interference occurs, why an interference pattern common to all signal photons should exist under any circumstances seems mysterious anyway, given the lack of such a common pattern for either the atoms or the electromagnetic field in the Scully-Englert-Walther experiment. In that experiment, linking the microwave cavities to let the microwave fields affect the photon detector between them does not result in a unique interference pattern for the atom. Any atom passing through the apparatus may with equal likelihood fit a constructive or a destructive interference pattern when the microwave cavities are linked, just as it does when the cavities are separated. The fact that signal beams in the Zou-Wang-Mandel experiment interfere at all under certain conditions thus appears about as problematic as the interference's obliteration through an apparently nonlocal mechanism.

With these puzzling features, some other explanation of the phenomenon seems called for. It turns out that a single state vector for the system can be proposed, consistent with the phenomena reported by Zou, Wang, and Mandel, that does not imply instantaneous actions at a distance at any time.

The state vector common to all the photon systems in each run of the experiment indicates no certain location for the original pump photon's down-conversion, but does indicate that either nothing meets the first down-converter, or the pump photon does. If the pump photon does meet this converter, it may induce a splitting of its energy into signal and idler beam photons. If the pump photon doesn't meet this converter, nothing happens there. Whereas either nothing or the pump photon meets the first down-converter, either the pump photon, or possibly an idler photon from the first down-converter's beam, meets the second down-converter. (An idler photon from the first converter's idler beam may or may not appear at the second converter if there is an attenuating barrier between the converters.) This is a significantly different condition. Whereas there is nothing to prevent down-conversion at the first converter from happening at any time, down-conversion at the second converter results in an idler that can interfere with the idler from the first converter. While at some times, the idler beam originating from the second converter will interfere constructively with that of the first, at other times the interference will be destructive, which means that down-conversion at the second converter will be inhibited then, which means that the second converter's *signal-photon production* will be inhibited then.

Following Zou, Wang, and Mandel, we consider the Hamiltonian for the action of the down-converters:

$$H_C = g_1 V_1(t) a^\dagger_{i1} a^\dagger_{s1} + g_2 V_2(t) a^\dagger_{i2} a^\dagger_{s2} \qquad (23)$$
$$+ g_1^* V_1^*(t) a_{i1} a_{s1} + g_2^* V_2^*(t) a_{i2} a_{s2},$$

$a_{i1}$, $a_{s1}$, $a_{i2}$, and $a_{s2}$ being annihilation operators for idler and signal photons at the first and second down-converters respectively, $V_1(t)$ and $V_2(t)$ being the amplitudes of the pump beam at the two down-converters at time t, and $g_1$ and $g_2$ being coupling constants for each down-converter. Account must also be taken of the attenuator between the two down-converters. Its action can be accounted for by a further interaction-Hamiltonian operator,

$$H_A = T a^\dagger_{i1} a_{i1} + B a_{i1} + B^* a^\dagger_{i1}, \qquad (24)$$

in which the first term represents the transmission of the idler photon through the attenuator, and the other terms represent its being blocked and absorbed and the inverse process (emission from the attenuator). (If the idler were deflected instead of absorbed, the Hamiltonian would be $T a^\dagger_{i1} a_{i1} + R a^\dagger_{i'1} a_{i1} + R^* a^\dagger_{i1} a_{i'1}$, the i´ indicating a different direction for the idler beam.)

Taking $H_A(t) H_C(t)$ as the interaction Hamiltonian for this experiment and $|0_{i1}, 0_{s1}, 0_{i2}, 0_{s2}\rangle$ to represent the state vector common to all runs of the experiment at a time when the pump beam is split but not yet down-converted, we get $1 + H_A(t')H_C(t)$ as a first approximation to a term of the S-matrix for the system, which means that the state vector becomes, at the end of an experimental run, approximately

$$|\Psi\rangle = |0_{i1}, 0_{s1}, 0_{i2}, 0_{s2}\rangle$$
$$+ g_1 V_1(t_1) (T|1_{i1}, 1_{s1}, 0_{i2}, 0_{s2}\rangle + B|0_{i1}, 1_{s1}, 0_{i2}, 0_{s2}\rangle) \quad (25)$$
$$+ g_2 V_2(t_2)|0_{i1}, 0_{s1}, 1_{i2}, 1_{s2}\rangle,$$

in which idler photons from each down-converter are only directly associated with signal photons from the same converter. Here "$1_x$" and "$0_x$" represent the presence and absence respectively of a photon in beam mode x.

One other detail remains to be accounted for. If the down-converters are properly aligned, the first and second down-converters' idler beams are identical except possibly for their phases. This possible phase difference is the reason down-conversion at the second converter is periodically inhibited, so that it is less likely at some times than others. If the phase factor of an idler produced at the first down-converter is taken to be unity at the time of production, that idler's phase at the second down-converter at time $t_2$ is

$$\omega_i[(t_2-t_1)-(d_{12}/c)], \quad (26)$$

$d_{12}$ being the distance between the down-converters and $\omega_i$ being the angular frequency of the idler beam. We may thus describe the state vector by

State Vectors and Physical States    70

$$|\Psi\rangle = |0_{i1}, 0_{s1}, 0_{i2}, 0_{s2}\rangle$$
$$+ g_1 V_1(t_1) \exp(\omega_i[(t_2-t_1)-d_{12}/c]) \quad (27)$$
$$\times (T|1_i, 1_{s1}, 0_{s2}\rangle + B|0_i, 1_{s1}, 0_{s2}\rangle)$$
$$+ g_2 V_2(t_2)|1_i, 0_{s1}, 1_{s2}\rangle,$$

representing the presence of an idler photon beyond the second down-converter by $|1_i\rangle$. We note here the phase difference just mentioned, observing that the second and fourth terms of equation (27) may add together or cancel depending on the time difference $t_2-t_1$ between down-conversions, and also note that the degree of this interference depends on the magnitude of T (whose square is proportional to the likelihood that a photon of the first down-converter's idler beam exists beyond the attenuator near the second down-converter).

Although this state vector is different from the one used by Zou, Wang, and Mandel, it implies the same dependence of the signal-interference's visibility on the attenuation of the first idler beam. Following their procedure and using the operator $E_s^+ = a_{s1}e^{i\theta_1} + a_{s2}e^{i\theta_2}$ to describe the electric field at the signal detector, with $\theta_1$ and $\theta_2$ being the phase shifts in the signals between the respective down-converters and the detector, we find

$$|g_1 V_1(t_1)|^2 + |g_2 V_2(t_2)|^2$$
$$+ 2|g_1^* V_1^*(t_1) g_2 V_2(t_2)||T| \quad (28)$$
$$\times \cos(\theta_2-\theta_1-\omega_i[(t_2-t_1)-d_{12}/c]$$
$$+\arg(g_2)-\arg(g_1)+\arg(V_2)-\arg(V_1)-\arg(T))$$

for the average electric field $\langle\psi|E_s^- E_s^+|\psi\rangle$ at the signal detector. The interference term is seen to vary with the down-conversions' time difference, as noted before, while the interference's visibility is found to be

$$2|g_1^* V_1^*(t_1) g_2 V_2(t_2)||T|/(|g_1 V_1(t_1)|^2 + |g_2 V_2(t_2)|^2), \qquad (29)$$

the same (though expressed here in slightly different notation) as what was found by Zou, Wang, and Mandel. But in this case, in keeping with the interpretation of quantum theory presented heretofore, by beginning with the premise that physical processes do not involve action at a distance we show both a possible reason for the interference pattern's existence under certain conditions and how that interference pattern depends on the attenuation of the first idler beam without assuming nonlocal influences. The influences assumed are purely local: if the first idler beam reaches the second down-converter, any time at which the formation of a second idler beam would put it out of phase with the first is a time when not only the second idler's formation, but a second signal-beam's formation, is inhibited, thus making a particular phase relationship between the two signal beams maximally probable and the orthogonal relation between signal-beam phases maximally improbable.

The search for such a "local-influence" way of explaining these phenomena has been prompted by the assumption that nonlocal influences, or instantaneous propagations of local ones, are inconsistent with quantum physics. Furthermore, the different mathematical premises of the explanations have different experimental consequences: the possibility or lack thereof that down-conversion of a single photon can produce signal and idler photons at different down-converters. The "local-influence" explanation also allows immediate deduction of an experimental consequence: that there should be a specific time lag from when the opacity or reflectivity of the idler-beam attenuator is changed to when a change is apparent in the signal-beam interference pattern.  For a change in the probability of down-conversion at the second converter depends on the strength of the first idler beam there, and it will take as long for this strength to change as it takes for the idler to propagate there from the attenuator; it also takes time for the strength of the second signal-beam component to change at the signal detector.  Thus the total time lag should be no less than the sum of the propagation times from the attenuator into the second down-converter and along the second signal-beam path thence into the signal detector.

## 10.  RECAPITULATION.  KEY HYPOTHESES.  OTHER PHENOMENA TO BE CONSIDERED

The preceding examples and discussion illustrate considerations that, taken together, constitute a way of thinking about quantum phenomena that differs from currently familiar interpretations.  The discussions of each of these physical processes and their phenomena point out a variety of implications of this way of thinking, particularly of the hypotheses that physical causes do not involve simultaneous action at a distance throughout a physical system, that physical systems' states have a one-to-many correspondence with state vectors, and that each state vector, which corresponds to one aspect of a system's state, must be consistent with the other state vectors that each correspond to other aspects of that state.  In the following, other features of the universe and its physical processes, which are of particular significance to us for how they epitomize additional aspects of physics' quantum character, are also considered in terms of this way of thinking.

## 11. THE TIME-DEVELOPMENT OF PARTICLE INTERACTIONS

As we have seen, the concept of concurrently-existing state vectors provides a new way of looking at the time-development of physical systems generally. In this section we consider some particular implications of these ideas related to how elementary particles interact.

Much of our understanding of the interactions affecting all physical processes is based on observing their effects on elementary-particle decays and collisions. To test our theories of these interactions and of the particles they affect, we calculate how probable particular observations are according to those theories and see if actual decay and collision results have the implied frequencies of occurrence. To estimate the amplitudes of these probabilities, one may take state vectors corresponding to one's knowledge of the elementary-particle system's initial state, multiply by an operator that according to theory describes (at least approximately) how the particles may interact, and multiply again by a state vector corresponding to the final observation.

We have been accustomed to think of two-particle collisions along the following lines. Two particles, each with definite momentum (or possibly with relatively definite position, though initial momenta are the parameters usually defined for collision calculations), approach each other. If the particles interact at all, they will absorb and/or emit various other particles, which may themselves during the collision absorb and/or emit still other particles, which may also absorb and/or emit other particles, &$c$. The collision ends when all the particle absorption and emission is done, leaving the final products, whether they be two particles of the original type with (most likely) different momenta, or other particles entirely. According to this point of view, the system's initial state, described by one state vector, changes during the collision, being transformed by the interactions to a different state described by another state vector. Elementary-particle decays may be thought of in a similar way, except that before the interactions there is only one particle, and any and all particle emissions and/or absorptions finally result in other particles.

A different way of thinking about such processes involves the idea that multiple state vectors can together describe one physical state. This mode of thought is perhaps most easily illustrated and explained here by recalling the previously discussed variation of the "Schrödinger's cat" experiment. Because of how the unstable nucleus was set to either trigger a mechanism which would change the spin-orientation of the other particle along its x-axis (if that nucleus decayed within a certain time interval) or leave the mechanism and the other particle's spin alone (if the decay didn't occur), the state vector initially characterizable by

$$|+x, \text{still-intact radionucleus}\rangle \quad (30)$$

has a final form

$$(|+x, \text{still-intact radionucleus}\rangle \quad (31)$$
$$+ e^{+i\phi}|-x, \text{decayed radionucleus}\rangle)/\sqrt{2}.$$

This final form is the sum of two state vectors, only one of which represents what we would observe at the end if we were to check the x-orientation of the spin and the integrity of the erstwhile (and possibly still) unstable nucleus. Both

$$|+x, \text{still-intact radionucleus}\rangle$$

and

$$(|+x, \text{still-intact radionucleus}\rangle$$
$$+ e^{+i\phi}|-x, \text{decayed radionucleus}\rangle)/\sqrt{2}$$

represent the same state vector, each expression simply representing it at different times.

As was stated earlier, the operator S describes the effect of all physical processes affecting the outcome of the experiment; there are no physical influences that could cause any collapse of the state vector to either $|+x, \text{still-intact radionucleus}\rangle$ or $e^{+i\phi}|-x, \text{decayed radionucleus}\rangle$. But if different, compatible state vectors exist, which together completely describe the system's physical state and individually describe different aspects or features of it according to the ideas heretofore presented and discussed, then either $|+x, \text{still-intact radionucleus}\rangle$ or $|-x, \text{decayed radionucleus}\rangle$ will properly describe a different aspect of this system--namely, the one we observe when we look for the intactness or decay of the nucleus and the spin-orientation along the x-axis. In terms of this point of view, the scalar products

$$\langle +x, \text{still-intact radionucleus}|S|+x, \text{still-intact radionucleus}\rangle \qquad (32)$$

and

$$\langle -x, \text{decayed radionucleus}|S|+x, \text{still-intact radionucleus}\rangle \qquad (33)$$

are not respective probability amplitudes for the x-orientation and nuclear coherence to *stay the same or change*, but are amplitudes that either

$$\langle +x, \text{still-intact radionucleus}| \tag{34}$$

or

$$\langle -x, \text{decayed radionucleus}| \tag{35}$$

*concurrently describe the system already known to be described* by

$$(|+x, \text{still-intact radionucleus}\rangle \tag{31´}$$
$$+ e^{+i\phi}|-x, \text{decayed radionucleus}\rangle)/\sqrt{2}.$$

Seen another way, in terms of the "pre-interaction" forms of $\langle +x, \text{still-intact radionucleus}|$ and $\langle -x, \text{decayed radionucleus}|$,

$$\langle +x, \text{still-intact radionucleus}|S|+x, \text{still-intact radionucleus}\rangle$$

and

$$\langle -x, \text{decayed radionucleus}|S|+x, \text{still-intact radionucleus}\rangle$$

represent the amplitudes that

$$(\langle +x, \text{still-intact radionucleus}| \tag{36}$$
$$- \langle -x, \text{decayed radionucleus}|e^{-i\phi})/\sqrt{2}$$

and

$$(\langle -x, \text{decayed radionucleus}| \qquad (37)$$

$$+ \langle +x, \text{still-intact radionucleus}|e^{+i\phi})/\sqrt{2}$$

coexist with $|+x, \text{still-intact radionucleus}\rangle$.

According to this way of thinking, what happens when we do an experiment with elementary particle decays or collisions could be described as follows. We set up the system we wish to observe so that initially it can be described--in *one* of its aspects--by some particular state vector, say $|p_1, p_2\rangle$ for a collision experiment with two particles of different momenta. The system is then affected by interactions of some kind whose features are described by an appropriate operator S. How these interactions affect the aspect of the system we described by $|p_1, p_2\rangle$ is expressed by the product $S|p_1, p_2\rangle$, which in the general case is not an eigenvector of momenta, but a sum of such momentum eigenvectors.

Depending on what interactions are involved, $S|p_1, p_2\rangle$ may not even be an eigenvector of any particle-number operators; the sum may consist of terms involving many different numbers and types of particles. This sum, whose exact form depends on what S is as well as on $|p_1, p_2\rangle$, represents the same state vector as $|p_1, p_2\rangle$ itself, at a later time. As such, it represents a feature of the system--the same feature that $|p_1, p_2\rangle$ represents. But in the usual sort of particle-collision experiments, it would not represent the aspect or kind of feature that we would observe after the collision. In essence, we normally look for eigenvectors of operators other than those that $S|p_1, p_2\rangle$ itself is an eigenvector of. That is, we are examining a different feature of the system than the one defined by $|p_1, p_2\rangle$--even if what we are looking for after the collision happens to be characterized by the same kinds of variables that describe the form of $|p_1, p_2\rangle$ before the collision.

By the same reasoning, whatever state vector we do observe after the collision--say, $|p_a, p_b, p_c\rangle$, an eigenvector of the momenta of three particles a, b, and c--will correspond to a *pre*-collision state vector that is also not in the general case an eigenvector of momentum or particle number.

If state vectors correspond to particular aspects of states instead of to states themselves, the interaction does not change the state of the system (and neither does observing the system), but the interaction changes each of the state vectors that describe different aspects of the system, in that each is an eigenvector of some particular set of operators--not necessarily the same set as for the other vectors--before the collision, and a different set of operators afterwards.

What, then, does the amplitude $\langle p_a, p_b, p_c|S|p_1, p_2\rangle$ represent? According to the ideas just described, it is the amplitude for the state vector $\langle p_a, p_b, p_c|$ to coexist with the vector $S|p_1, p_2\rangle$--which means exactly the same thing as the amplitude that $\langle p_a, p_b, p_c|S$ coexists with $|p_1, p_2\rangle$. Since "$\langle \Psi|$" is basically just another way of expressing "$|\Psi\rangle$", we could equally well say that $\langle p_a, p_b, p_c|S|p_1, p_2\rangle$ is the amplitude that $|p_a, p_b, p_c\rangle$ coexists with $S|p_1, p_2\rangle$, or that $\langle p_a, p_b, p_c|$ and $\langle p_1, p_2|S^\dagger$ both concurrently describe the same aspect of the colliding-particle system's state, or that this state has different features respectively described by $S^\dagger|p_a, p_b, p_c\rangle$ and $|p_1, p_2\rangle$. (One does have to take care that one doesn't say the amplitude $\langle p_a, p_b, p_c|S|p_1, p_2\rangle$ equals $\langle p_1, p_2|S^\dagger|p_a, p_b, p_c\rangle$, since it actually equals the complex conjugate of $\langle p_1, p_2|S^\dagger|p_a, p_b, p_c\rangle$, or $\langle p_1, p_2|S^\dagger|p_a, p_b, p_c\rangle^*$. However, both these amplitudes of course imply the same probability of coexistence, namely $\langle p_1, p_2|S^\dagger|p_a, p_b, p_c\rangle\langle p_a, p_b, p_c|S|p_1, p_2\rangle$.) According to this view, Feynman diagrams should be seen as representing, not processes whereby an initial state described

by vector |A> becomes a state described by |B>, or even processes that contribute to state |A>'s transformation into state |B>, but single states which are characterized by both state vectors |A> and |B> (along with still other vectors representing other aspects not accounted for in the diagrams).

Interestingly, one could represent $<p_a, p_b, p_c|S|p_1, p_2>$ by $<p_a, p_b, p_c|(S^\dagger S)S|p_1, p_2> = (<p_a, p_b, p_c|S^\dagger)S(S|p_1, p_2>)$, which corresponds to a rather different-looking Feynman diagram. The diagrams for $<p_a, p_b, p_c|S|p_1, p_2>$ and $(<p_a, p_b, p_c|S^\dagger)S(S|p_1, p_2>)$ must therefore be equivalent; the difference between them is that the diagram for $(<p_a, p_b, p_c|S^\dagger)S(S|p_1, p_2>)$ represents the same state vectors by the forms they take at times other than when we would normally observe them, when they may not even be eigenvectors of the observables we are interested in or would recognize them by.

## 12. PERCEPTION AND STATE VECTORS

The fact that state vectors at certain times may take forms by which we would not recognize or readily identify them, at least under certain conditions, brings up a significant point about how we do determine whether a particular state vector characterizes a system's state. In this section, we first examine another familiar physical process in terms of the idea that any physical state is represented by a set of many state vectors. Afterward we will consider how measuring and perceiving only one aspect of a state at a time may relate to our tendency to equate state vectors with states themselves rather than single aspects of states.

One feature of the photoelectric effect that appears surprising according to classical theory is the fact that the energy of the electrons liberated from illuminated metal shows no dependence on the illumination's intensity. This is no mystery in terms of quantum theory; the energy of each photoelectron is the energy of one absorbed photon minus whatever energy the electron expends getting out of the metal. Intense light simply means more photons (and more photoelectrons), not more energy per liberated electron. A similarly surprising feature is that illuminating a metal surface liberates electrons immediately, regardless of how dim the light is. From classical electrodynamics, one would expect that it would take some time for an electromagnetic wave to drive an electron into oscillations of higher and higher amplitude until it is shaken free of the positive charge in the metal. Again, quantum theory provides a resolution: there is some nonzero probability that one photon is present anywhere the metal surface is illuminated, no matter how little illumination there is per unit area of metal on the whole; wherever the photon does happen to show up and get absorbed, there may an electron be emitted (if the photon's energy is great enough).

But according to the concept that many state vectors together describe or determine a system's state, the illumination-and-metal system is not only described by a state vector for a well-localized photon energizing an electron (while being absorbed), but by another vector--a vector for the electromagnetic field spread so evenly over the metal surface that the classical energy flux is too low to shake an electron loose as rapidly as we know can happen. (If we experimented with only one photon and quit, we could not infer this state vector's existence; we determine it either by repeated experiments, which can reveal that only this particular state vector can characterize all the photons directed at the metal--much as we could determine that all three-spin systems we observe in the Greenberger, Horne, and Zeilinger experiment were characterized by (|+z, +z, +z⟩ - |-z, -z, -z⟩)/√2--or by examining the light source and seeing that it is always characterized by a state vector consistent with an even illumination of the metal.) As long as we think photons are "really" localized, at least at the moment electrons absorb their energy, it does not seem so strange that even dim light can produce photoelectrons. But if the dimness of evenly-spread light is characterized by a state vector of its own, how can the photoelectric effect happen? How does a physical state characterized by |"low, even illumination"⟩ come to fit |"one photoelectron emitted from location x"⟩ in the same short time, regardless of how low the illumination is?

By now the answer may be obvious--the state vector for |"low illumination"⟩ does not become the state vector for |"one well-localized photoelectron"⟩; it becomes something else, and the "localized photoelectron" state vector is the final form of a different state vector that started out as something other than the one for low, even illumination. We could demonstrate this by using the full resources of standard quantum electrodynamical operators, which can be used to describe any light waves of all possible frequencies, intensities, polarizations, and so forth, along with all manner of electron configurations in any sort of metal. However, the essential aspects can be described and analyzed much more simply.

Consider a very simple piece of metal made up of two "metallic ions" bound by two "free electrons" (free within the metal). This piece of metal has only two distinct locations of any significance: +x and -x. If a single photon enters either of these places, it may be absorbed and leave its energy to immediately produce a photoelectron there. Alternatively, something different may happen instead--the photon could be reflected, or pass through the metal, or do something else. We can represent these possibilities by the following S operator:

$$(A + B_{+x}b_{+x}^{\dagger}a_{+x} + C_{+x}c_{+x}^{\dagger}a_{+x} - B_{+x}^{*}b_{+x}a_{+x}^{\dagger} - C_{+x}^{*}c_{+x}a_{+x}^{\dagger}$$
$$+ B_{-x}b_{-x}^{\dagger}a_{-x} + C_{-x}c_{-x}^{\dagger}a_{-x} - B_{-x}^{*}b_{-x}a_{-x}^{\dagger} - C_{-x}^{*}c_{-x}a_{-x}^{\dagger}) \qquad (38)$$
$$\times 1/(A^{*}A + B_{+x}^{*}B_{+x} + C_{+x}^{*}C_{+x} + B_{-x}^{*}B_{-x} + C_{-x}^{*}C_{-x}).$$

Here, the $B_{\pm x}$ and $C_{\pm x}$ are constants and the $B_{\pm x}^{*}$ and $C_{\pm x}^{*}$ are their respective complex conjugates. The operators $a_{+x}$ and $a_{-x}$ represent absorption of a photon at +x or -x respectively, while their Hermitian conjugates represent photon emission at those locations. $b_{+x}^{\dagger}$ and $b_{-x}^{\dagger}$ represent photoelectron emission at +x and -x, while their conjugates represent electron absorption. The $c^{\dagger}$s are operators for production of anything other than single electrons or photons; the cs stand for their inverses. A is a constant, representing relative nondisturbance of photons, electrons, or anything else by the metal.

For the operators $a_{\pm x}$, $b_{\pm x}$, $c_{\pm x}$ and their conjugates, the following rules apply (with o a generic operator representing either a, b, or c):

$o_{\pm x}|1_{\pm x}\rangle = |0_{\pm x}\rangle$ (absorption of a single photon (if o=a), electron (if o=b), or other things or combinations of things that might be produced during photoabsorption (if o=c) at location ±x, leaving no photon/electron/other entity there);

$o_{\pm x}^{\dagger}|0_{\pm x}\rangle = |1_{\pm x}\rangle$ (emission of a photon, photoelectron, or other entity produced where there was none before); (39)

$o_{\pm x}|0_{\pm x}\rangle = 0$;

$b_{\pm x}^{\dagger}|1_{\pm x}\rangle = 0$ ($b_{\pm x}^{\dagger}$, since false for $a_{\pm x}^{\dagger}$, not significant for $c_{\pm x}^{\dagger}$);

$o_{\pm x}|0_{\mp x}\rangle = o_{\pm x}^{\dagger}|1_{\mp x}\rangle = 0$.

If the light is so dim that only one photon at a time illuminates the metal, all the possible state vectors may be expressed in the form $|I_{+x}, I_{-x}, J_{+x}, J_{-x}, K_{+x}, K_{-x}\rangle$, in which the I, J, and K are either 0 or 1, with the I representing the number of photons at +x or -x, the J being the number of electrons at either place respectively (being absorbed, photoemitted, or left alone as the case may be), and the K representing the presence or absence of "other entities" that may be involved during photon interactions with the metal (or undisturbed otherwise).

The presence of an electron free of the metal at +x (and nothing more, except the metal itself) is thus represented by the state vector

$$|0_{+x}, 0_{-x}, 1_{+x}, 0_{-x}, 0_{+x}, 0_{-x}\rangle \tag{40}$$

while the presence of a similar electron (and nothing else) at -x is represented by

$$|0_{+x}, 0_{-x}, 0_{+x}, 1_{-x}, 0_{+x}, 0_{-x}\rangle. \tag{41}$$

Likewise, the state vector $|1_{+x}, 0_{-x}, 0_{+x}, 0_{-x}, 0_{+x}, 0_{-x}\rangle$ represents the presence of one photon at location +x on the metal, while $|0_{+x}, 1_{-x}, 0_{+x}, 0_{-x}, 0_{+x}, 0_{-x}\rangle$ represents a photon's presence at location -x. Dim light, containing no more than one photon, that "evenly" illuminates the metal, can thus be represented by

$$\begin{aligned}(e^{+i\phi}|1_{+x}, 0_{-x}, 0_{+x}, 0_{-x}, 0_{+x}, 0_{-x}\rangle \\ + e^{-i\phi}|0_{+x}, 1_{-x}, 0_{+x}, 0_{-x}, 0_{+x}, 0_{-x}\rangle)/\sqrt{2}.\end{aligned} \tag{42}$$

$S(e^{+i\phi}|1_{+x}, 0_{-x}, 0_{+x}, 0_{-x}, 0_{+x}, 0_{-x}\rangle + e^{-i\phi}|0_{+x}, 1_{-x}, 0_{+x}, 0_{-x}, 0_{+x}, 0_{-x}\rangle)/\sqrt{2}$, the form of this state vector after the photon has interacted with (or passed through) the metal, is equal to

$$\begin{aligned}(Ae^{+i\phi}&|1_{+x}, 0_{-x}, 0_{+x}, 0_{-x}, 0_{+x}, 0_{-x}\rangle \\ + Ae^{-i\phi}&|0_{+x}, 1_{-x}, 0_{+x}, 0_{-x}, 0_{+x}, 0_{-x}\rangle \\ + B_{+x}e^{+i\phi}&|0_{+x}, 0_{-x}, 1_{+x}, 0_{-x}, 0_{+x}, 0_{-x}\rangle \\ + B_{-x}e^{-i\phi}&|0_{+x}, 0_{-x}, 0_{+x}, 1_{-x}, 0_{+x}, 0_{-x}\rangle \\ + C_{+x}e^{+i\phi}&|0_{+x}, 0_{-x}, 0_{+x}, 0_{-x}, 1_{+x}, 0_{-x}\rangle \\ + C_{-x}e^{-i\phi}&|0_{+x}, 0_{-x}, 0_{+x}, 0_{-x}, 0_{+x}, 1_{-x}\rangle) \\ \times\, 1/(&\sqrt{2}(A^*A + B_{+x}^*B_{+x} + C_{+x}^*C_{+x} + B_{-x}^*B_{-x} + C_{-x}^*C_{-x})).\end{aligned} \tag{43}$$

We see that, if the metal is evenly illuminated, the light may pass through the metal undisturbed, or it may be absorbed and generate something other than a photoelectron, or it may be absorbed and produce a photoelectron. If it does produce a photoelectron, the photoelectron is equally likely to come from location +x or location -x. This state vector--the post-interaction version or form of the pre-interaction form $(e^{+i\phi}|1_{+x}, 0_{-x}, 0_{+x}, 0_{-x}, 0_{+x}, 0_{-x}\rangle + e^{-i\phi}|0_{+x}, 1_{-x}, 0_{+x}, 0_{-x}, 0_{+x}, 0_{-x}\rangle)/\sqrt{2}$--is clearly *compatible* with either of the post-interaction "one localized photoelectron" vectors $|0_{+x}, 0_{-x}, 1_{+x}, 0_{-x}, 0_{+x}, 0_{-x}\rangle$ and $|0_{+x}, 0_{-x}, 0_{+x}, 1_{-x}, 0_{+x}, 0_{-x}\rangle$, but it is not *itself* a |*"low-illumination"*, "one localized photoelectron"⟩ vector.

(As a point of interest, we can also easily calculate the "pre-interaction" version of both "local electron" state vectors by determining the products $\langle 0_{+x}, 0_{-x}, 1_{+x}, 0_{-x}, 0_{+x}, 0_{-x}|S$ and $\langle 0_{+x}, 0_{-x}, 0_{+x}, 1_{-x}, 0_{+x}, 0_{-x}|S$. These are

$$(\langle 0_{+x}, 0_{-x}, 1_{+x}, 0_{-x}, 0_{+x}, 0_{-x}|A$$
$$+ \langle 1_{+x}, 0_{-x}, 0_{+x}, 0_{-x}, 0_{+x}, 0_{-x}|B_{+x})  \quad (44)$$
$$\times 1/(A^*A + B_{+x}{}^*B_{+x} + C_{+x}{}^*C_{+x} + B_{-x}{}^*B_{-x} + C_{-x}{}^*C_{-x})$$

and

$$(\langle 0_{+x}, 0_{-x}, 0_{+x}, 1_{-x}, 0_{+x}, 0_{-x}|A$$
$$+ \langle 0_{+x}, 1_{-x}, 0_{+x}, 0_{-x}, 0_{+x}, 0_{-x}|B_{-x}) \quad (45)$$
$$\times 1/(A^*A + B_{+x}{}^*B_{+x} + C_{+x}{}^*C_{+x} + B_{-x}{}^*B_{-x} + C_{-x}{}^*C_{-x}),$$

or (in ket form)

$$(A^*|0_{+x}, 0_{-x}, 1_{+x}, 0_{-x}, 0_{+x}, 0_{-x}\rangle$$
$$+ B_{+x}{}^*|1_{+x}, 0_{-x}, 0_{+x}, 0_{-x}, 0_{+x}, 0_{-x}\rangle) \qquad (46)$$
$$\times 1/(A^*A + B_{+x}{}^*B_{+x} + C_{+x}{}^*C_{+x} + B_{-x}{}^*B_{-x} + C_{-x}{}^*C_{-x})$$

and

$$(A^*|0_{+x}, 0_{-x}, 0_{+x}, 1_{-x}, 0_{+x}, 0_{-x}\rangle$$
$$+ B_{-x}{}^*|0_{+x}, 1_{-x}, 0_{+x}, 0_{-x}, 0_{+x}, 0_{-x}\rangle) \qquad (47)$$
$$\times 1/(A^*A + B_{+x}{}^*B_{+x} + C_{+x}{}^*C_{+x} + B_{-x}{}^*B_{-x} + C_{-x}{}^*C_{-x})$$

respectively. This shows that the "pre-interaction" form of both vectors is in general a combination of one state vector for a localized photon that could liberate the electron into its localized final state, and another state vector representing the presence of an electron at the same location already free of the metal with no photon there at that moment.)

As with the preceding analysis of the many-state-vector-per-state interpretation of elementary-particle interactions, we see that the alternative forms of both "local particle" state vectors--the "pre-interaction" form of the local-electron vectors and the "post-interaction" form of the even-illumination vector--might not correspond to anything we would easily perceive directly. At first sight, it appears that the uniform, "nonlocal" illumination of the metal results in something definitely nonuniform. On closer examination, we see that the uniformity and nonuniformity relate to rather more separate things, and that the uniform one stays uniform while the nonuniform one stays nonuniform. The state vector whose initial condition represents the uniform illumination of both regions of the metal develops into one that represents propagating light, a propagating electron, and other sorts of disturbance, which are all associated as much with one region of the metal as with the other. The other state vector, whose final form represents a locatable electron, has an initial form equally associated with the same location (either the +x or the -x region, depending on the system's state). The predecessor of a local electron may be the same local electron or a local (not uniform-illumination) photon, in the sense that the local aspect of the system's final condition is a direct consequence of an equally local aspect of its pre-interaction condition. The local aspect of this system's state may be ill-defined with respect to what *kind* of local particle comprises it before the interaction occurs, but what location the particle has, whatever the particle is, is always quite definite--equally definite before and after the interaction.

A simpler system of this type, in which photons are inevitably absorbed to produce photoelectrons, is more easily analyzed. In this case, A and the Cs all equal zero, so the final form of

$$(e^{+i\phi}|1_{+x}, 0_{-x}, 0_{+x}, 0_{-x}, 0_{+x}, 0_{-x}\rangle \qquad (48)$$
$$+ e^{-i\phi}|0_{+x}, 1_{-x}, 0_{+x}, 0_{-x}, 0_{+x}, 0_{-x}\rangle)/\sqrt{2}$$

is

$$(B_{+x}e^{+i\phi}|0_{+x}, 0_{-x}, 1_{+x}, 0_{-x}, 0_{+x}, 0_{-x}\rangle$$
$$+ B_{-x}e^{-i\phi}|0_{+x}, 0_{-x}, 0_{+x}, 1_{-x}, 0_{+x}, 0_{-x}\rangle) \qquad (49)$$
$$\times 1/(\sqrt{2}(B_{+x}{}^*B_{+x} + B_{-x}{}^*B_{-x}))$$

while the initial forms of

$$|0_{+x}, 0_{-x}, 1_{+x}, 0_{-x}, 0_{+x}, 0_{-x}\rangle \qquad (50)$$

and

$$|0_{+x}, 0_{-x}, 0_{+x}, 1_{-x}, 0_{+x}, 0_{-x}\rangle \quad (51)$$

are

$$(B_{+x}{}^*|1_{+x}, 0_{-x}, 0_{+x}, 0_{-x}, 0_{+x}, 0_{-x}\rangle)/(B_{+x}{}^*B_{+x} + B_{-x}{}^*B_{-x}) \quad (52)$$

and

$$(B_{+x}{}^*|0_{+x}, 1_{-x}, 0_{+x}, 0_{-x}, 0_{+x}, 0_{-x}\rangle)/(B_{+x}{}^*B_{+x} + B_{-x}{}^*B_{-x}). \quad (53)$$

For this type of system, we see that state vectors representing a state with initially localized photons equivalently represent states with finally localized photoelectron emissions, while state vectors that represent even, dim illumination of the metal also imply equally even, low-flux photoelectron emission.

    We are not accustomed to perceiving, or discussing, photoelectron emission in those terms. We could measure the photoelectron emissions over a period of time in which the metal surface underwent low-intensity illumination, and determine from the frequency of photoelectron emission from both the +x and the -x locations on the metal that the photoemission was uniform overall. The photoelectron emission itself is evidence--the kind of evidence we often take--that the illumination of the metal is equal for the +x and -x locations.

    According to the multiple-state-vectors-per-state concept, uniform illumination does not "result" in a local photoelectron, in the sense that the time development of a uniform-illumination *state vector* does not end with its becoming a state vector for local

photoelectron emission. Local photoelectron emission is represented by a state vector whose pre-interaction version represents a local photon; the post-interaction version of a uniform-illumination vector is a vector for "uniform photoelectron emission", something we do not usually observe directly. The aspects of this system that we might take direct notice of--the uniform illumination and the local photoelectron emission--have to be represented by two different time-dependent state vectors, not one.

This idea is a key point for understanding how single states could be represented by multiple state vectors. Though a multiplicity of state vectors may each represent just one physical feature of a physical state, experience shows that we only measure one aspect of a given system at a time. If we measure a system to determine the value of any one of its parameters, we will only have ascertained something about those state vectors that specify that parameter; other state vectors specify values of different parameters altogether, specifying none for the parameters we measure.

A point of interest here is that, when we concentrate on measuring locations of things in physical systems, we do not directly perceive other system aspects that are not usually noticed; indeed, we are accustomed to seeing mainly the "location" aspects of physical states, to the point that we infer other aspects of states (such as momenta, spins, energies, &$c$.) mainly in terms of observations of location aspects. We may have a tendency to regard only state vectors related to "location aspects" as observable or "real"; if so, our way of observing other aspects of physical systems through location aspects may have much to do with this tendency. Indeed, expressing many state vectors as wavefunctions in the position representation can help us visualize them; wavefunctions of this type have a fairly direct correspondence to how we often visualize classical waves. But for any particular state of a set of interacting classical fields, there is only one time-varying wavefunction for each field. For the state vectors of interest in the photoelectric system, the position representation includes two different photon wavefunctions and two different photoelectron wavefunctions, only one each of which correspond to customary observations; furthermore, if photoelectron production is less than 100% certain, none of these wavefunctions (or pairings thereof) are different-time versions of others.

Our tendency to think of "the" state vector of a physical system in its position representation may say something about the ways we are able and unable to perceive physical systems.  The forms of colliding-particle state vectors are clearer and more familiar when represented on the side of the collision time when they correspond most nearly to a definite number of particles of definite types and momenta, which latter quantities can be determined by various sorts of position-related measurements on the ingoing and outgoing particles.  For some types of interaction, though, these same vectors' continuations beyond the collision event may approach eigenvectors of neither number nor momentum operators for distinct types of particle (though the *total* momentum of the system is just as definite for both state vectors in either their pre- or post-collision forms).

We are aware, however, of other things besides position *per se*.  The original form of the Schrödinger cat paradox involved, not a particle with spin, but a cat that would be killed if its air supply were contaminated with poison after the decaying radionucleus triggered the poison's release.  It takes relatively little imagination for one familiar with quantum theory to imagine a spin-and-radionucleus system characterized by a superposition of definite-spin, definite-nuclear-composition state vectors, especially as a superposition of spins along one axis is equivalent to a definite spin for that particle along a different axis.  It may seem a bit more problematic to think of how the cat could be characterized by a superposition of "live" and "dead" state vectors. (There are actually many different "live" and "dead" state vectors, since a live cat or a dead one might be in any number of physical states.) According to customary interpretations of quantum theory, it seems that the cat would have to be somehow alive and dead at the same time until it was observed. But on considering the preceding analysis of the spin-nucleus system, we may realize that, whatever state vector we *finally* observe for the cat, the form of that observation's state vector *before* the radionucleus is connected to the trigger is also a superposition of "live" and "dead" vectors. Just as "live cat + intact nucleus" state vectors become a superposition of "live cat + intact nucleus" and "dead cat + decayed nucleus" state vectors while the radionucleus is connected to the poison release, so would an originally-existing superposition become a "live cat" or a "dead cat" state vector during that same process--the final result depending on the originally-superposed vectors' relative phase. We (and the cat) may be well aware of the cat's being alive at the first. But we cannot (nor, presumably, can the cat) directly perceive what the superpositions are. We cannot perceive the original, pre-interaction linear combination of the "cat + nucleus" vector until time passes and it has become what it must--either a "live cat" or a "dead cat" vector. And we cannot directly perceive the final superposition of the "live cat" state vector with the "dead cat" state vector, though we can deduce what the superposition is mathematically. The point here is that, if any physical state is defined by many state vectors together, organisms are at all times characterized not only by such state vectors as specify their states of livelihood, but by other state vectors too, which are linear combinations of the preceding kind, that do not represent anything directly perceivable unless the system changes in a way that renders at least some of what they represent directly perceivable.

## 13. STATE VECTORS AND THE UNIVERSE AS A WHOLE

The proposition that a consistent set of many state vectors describes the actual state of any physical system, and thus determines the outcome of any experiment involving that system, has clear implications for questions about the "wavefunction (or state vector) of the universe": namely, that not even in principle could a single state vector characterize the universe as a whole--and not because of the universe's complexity. The universe would have to be characterized by many state vectors, each compatible with the others, none being orthogonal to the others (though any pairs might be as nearly orthogonal as possible up to the limit of orthogonality). The puzzle of what a state vector for the entire universe would mean, seeing that measurements of the universe take place within the system being measured--unlike the usual situation in which the measurement apparatus is outside the observed system (though there is no clear boundary between the two anyway)--does not exist if multiple state vectors are involved. In relation to any one state vector, other state vectors would be, in effect, nonlocal hidden variables, each state vector corresponding to a particular aspect of the universe and its constituents.

## 14. INTERPRETATION OF QUANTUM THEORY AND PHYSICAL INSIGHT

Physical laws determine the time development of individual state vectors, and the orthogonality or nonorthogonality of pairs of state vectors determine what state-vector combinations might simultaneously characterize any physical system. The idea that multiple state vectors characterize a system's state concurrently, not just sequentially, appears to resolve the paradoxes associated with quantum theory. For one of the aforementioned examples, the only differences between predictions based on this idea and the usual predictions about quantum phenomena are infinitesimal. But for another, use of the conceptual scheme associated with this idea leads to experimental predictions which, while consistent with quantum theory as ordinarily formulated, seem less obvious without its use.

It remains to be seen whether the concept that many concurrent state vectors define one state leads to many other new predictions. If this concept and the others associated with it lead to correct conclusions, their greater value may reside in providing a clearer insight into why quantum-theory calculations, whose physical meaning can otherwise seem counterintuitive, accurately predict real phenomena nonetheless--thus making the counterintuitive intuitive.

According to classical field theory, any field configuration is synonymous with a unique state of the field. A difference in the field, no matter how slight, means that the field is in a different state. All differing classical fields are thus "orthogonal"; while one may calculate how much one field shape resembles another according to certain criteria-- e.g., by considering how alike their Fourier spectra are--such resemblance between fields has no classical association with physically significant probabilities. The classical identification of different wavefunctions with different field states, if retained in our thinking as we try to consider quantum physics, can lead us into paradoxes since, if what we observe in a field is not a constant of its motion, the wavefunction (or state vector represented thereby) that fits the observed value now is, in the general case, not physically capable of developing into one for the value we observe later. Such occurrences, whose (nonzero) probabilities are mathematically predictable from quantum-theory principles, seem to conflict with other seemingly well-evidenced physical principles like causality and the relativity of simultaneity. Where such conflicts exist, intuition may become unreliable. One cannot then be sure of which principles to trust when one tries to imagine a physical process without detailed calculation. One may be guided by calculation, of course, but may lack a sense of why the calculation works, or of how the principles one ignored in calculation fit in with the result.

The idea that different wavefunctions (or the state vectors they represent) do not directly equate to different states appears not to lead to such conflicts between physical principles. No instantaneous transmission of information among dispersed measuring apparati or other objects need occur to account for how the original state vector collapses, if it doesn't collapse. This would also answer the question of how the transmission is instantaneous throughout the system, no matter what the frame of reference is. The puzzle of which stage in a chain of observations and observations of observations is the one in which the wavefunction changes into a different one--a puzzle related to the question of where the boundary is between observer and observed system--would also be resolved, if the state vector observed by the later measurement is a different one. Finally, it would not seem that inanimate systems have to be constantly making random choices about what to do next. Each state vector develops according to the same causal law. If many state vectors correspond to one state, each vector representing a different aspect of the state, the probabilities we find are probabilities that one state vector describes a system concurrently with another; these need not be probabilities that one vector changes into another acausally.

If these assumptions are correct, a reliable way to think through any quantum process, while still bringing one's knowledge of other physical concepts to bear on the analysis with confidence that all of them will work, is to think the process through, from beginning to end, for *all* the relevant state vectors, and not track just one state vector at a time between measurement events as though the state itself changed suddenly with each. For classical systems, one only need do this for one "state vector", since classically the probability of two distinct field-and-particle configurations coexisting, even at just the measuring apparati when observations are made, is zero. More than one state vector may be required to describe relevant features of a quantum system fully; if so, one must keep track of as many state vectors as are relevant to the prediction one wants to make--not only of how each one develops in time, but of whether they are consistent with each other, of whether observational outcomes implied by some state vectors and state-vector combinations conflict with those implied by others.

The recognition that state vectors and states may have many-to-one instead of one-to-one correspondences appears potentially helpful for better understanding quantum theory and conclusions drawn from it, seeing how quantum theory fits in with other principles of physics that seemingly conflicted with it, and arriving at our conclusions about physical phenomena more reliably with greater confidence, by removing doubts about how the known principles of physics fit together and about their range of applicability for quantum physics. Accounting for these principles in the analysis of quantum systems can guide one to new conclusions of physical consequence.

**ACKNOWLEDGEMENTS**

My thanks to Wolfgang Rindler, Cyrus D. Cantrell, Robert Rutkowski, and Larry Williams for their helpful discussions and suggestions regarding this presentation, and to Fabrice Meriaudeau for translating the abstract into French.

I also wish to thank one other person. Earlier, I had supposed that the Zou-Wang-Mandel experiment was no different in any important way from that proposed by Scully, Englert, and Walther. This misunderstanding was based on my interpretation of a secondary-source account of the experiment. While the source was accurate, it was not immediately clear to me that what the signal beam detector showed was not an incoherent overlay of interference patterns like that of the Scully-Englert-Walther experiment. A version of this paper submitted for publication earlier reflected this misunderstanding. An anonymous reviewer of that version mentioned that the source I cited for the Zou-Wang-Mandel experiment was not the best, so I sought a copy of the original paper. It was only then that I learned the authors had discovered that the signal beams' interference was correlated differently to the idler-beam setup from the way the interference of Scully, Englert, and Walther's atoms correlated with the setup of their microwave cavity. This understanding is obviously crucial to the discussion of the Zou-Wang-Mandel experiment presented above. My thanks, then, to that reviewer for mentioning the preferability of the primary source.

**Endnotes**

[1] This problem is discussed in Fritz London and Edmond Bauer's paper "The theory of observation in quantum mechanics" (Ref. 1), in which they propose a solution to the difficulty that differs from the one to be proposed here.

[2] Erwin Schrödinger's "cat" illustration is given in one paragraph of his three-part work "Die gegenwärtige Situation in der Quantenmechanik" (Ref. 7). Eugene P. Wigner's "Remarks on the Mind-Body Question" (Ref. 8) describes a number of thought experiments in which the effects of an observer and his friend (another observer) on observed systems' wave functions are considered--though none of Wigner's illustrations in this article specifically involve a cat.

[3] Since this demonstration reviews Bell's argument, and since conventionally the $\sigma_i$ represent 2x2 Pauli matrices and Bell did not specify otherwise, 2x2 matrices are assumed here. It may be objected that for consistency with relativity, one should use four-component spinors and 4x4 matrices for the spin instead of two-component spinors and 2x2 matrices. But if we use the Foldy-Wouthuysen representation for spin-1/2 particles, the 4x4 matrices for what Foldy and Wouthuysen call the mean spin are simply $[[\sigma, 0]^T, [0, \sigma]^T]$, and the roles of inner products $\sigma \cdot \mathbf{a}$ and $\sigma \cdot \mathbf{b}$ in the argument are filled by $[[\sigma \cdot \mathbf{a}, 0]^T, [0, \sigma \cdot \mathbf{a}]^T]$ and $[[\sigma \cdot \mathbf{b}, 0]^T, [0, \sigma \cdot \mathbf{b}]^T]$. Indeed, in Ref. 9 Foldy and Wouthuysen use the symbol $\sigma$ to represent the 4x4 matrix itself, so in their notation, $\sigma \cdot \mathbf{a}$ and $\sigma \cdot \mathbf{b}$ would actually stand for their 4x4 versions. One can replace $\sigma \cdot \mathbf{a}$ and $\sigma \cdot \mathbf{b}$ with their 4x4 mean-spin counterparts and interpret the |A, B> as eigenvectors of these 4x4 matrices instead, and the demonstration follows exactly as before.

[4] Methods of determining such wave functions by measurements of a single particle are discussed by Y. Aharonov, J. Anandan, and L. Vaidman in Ref. 10.

[5] Aharonov and Vaidman discuss a way of measuring a single particle which could determine two of its state vectors at once through measurements that each relate to both vectors (Ref. 11). However, in this paper they address a question in nonrelativistic quantum theory, and note that in general the method discussed involves nonlocal interactions which can contradict relativistic causality (p. 371).

[6] See p. 112 of Ref. 5. The lack of uncontrolled disturbance to the atom's momentum in a classical sense does not mean that momentum is not affected in a more subtle sense; see Ref. 12.

[7] In hindsight, it appears that our (seemingly common) assumption, that uncontrolled disturbance of the observed particle by whatever illuminates it is what destroys its interference pattern, may derive (fallaciously) from our familiarity with the Heisenberg microscope thought-experiment. This experiment actually demonstrates that the quantum behavior of light makes the uncertainty principle inevitable for anything illuminated by it--specifically, that the Planck relation $E=h\nu$ is *sufficient* to guarantee that the product of uncertainties in a particle's momentum and position along one axis is $h/4\pi$; the experiment does not show that uncontrolled disturbance is actually *necessary* for the uncertainty principle to be true.

[8] As Ref. 13 states, Daniel Greenberger, Michael Horne, and Anton Zeilinger invented a new version of the EPR experiment (Ref. 14) which was then analyzed by Robert Clifton, Michael Redhead, and Jeremy Butterfield (Ref. 15), whose analysis inspired Mermin's.

[9] As in the discussion of Bell's thought-experiment, we may interpret the $\sigma_1 \cdot \mathbf{x}$ and $\sigma_1 \cdot \mathbf{y}$ as 4x4 matrices related to Foldy-Wouthuysen mean spin instead of as 2x2 spin matrices.

[10] The corresponding eigenspinors for $[[\sigma \cdot \mathbf{n}_i, 0]^T, [0, \sigma \cdot \mathbf{n}_1]^T]$ are

$(1/\sqrt{2})[\sin \theta_i/\sqrt{(1-\cos \theta_i)}, \exp(+i\phi_i)\cdot\sqrt{(1-\cos \theta_i)}, 0, 0]^T$ and

$(1/\sqrt{2})[0, 0, \sin \theta_i/\sqrt{(1-\cos \theta_i)}, \exp(+i\phi_i)\cdot\sqrt{(1-\cos \theta_i)}]^T$, one describing a particle and the other an antiparticle. Thus the argument takes the same form and leads to the same conclusions when we account for relativity explicitly by using four-component spinors. By way of definiteness, the illustration assumes we are dealing with particles instead of antiparticles, so we can simplify the notation by ignoring the extra two components.

[11] Interestingly, this recalls an argument of Bell's for not specifying hidden-variable values corresponding to similar special cases for the two-particle system he discussed. See Ref. 16.

[12] The authors themselves do not expect an idler photon from the first down-converter and a signal photon from the second down-converter to coincide, as they make clear on p. 320 of Ref. 17. The implication that a single down-conversion may produce signal and idler photons in different places is a logical consequence of combining their equations (1) and (2).

**William N. Watson**

Office of Scientific and Technical Information

United States Department of Energy

P. O. Box 62

Oak Ridge, TN  37831